\documentclass[aps,twocolumn,amssymb,showpacs]{revtex4}
\usepackage{graphicx}

\begin{document}

\title{Model Hamiltonian for Topological Insulators}

\author{Chao-Xing Liu$^{1}$, Xiao-Liang Qi$^2$, HaiJun Zhang$^3$, Xi Dai$^3$, Zhong Fang$^3$ and Shou-Cheng Zhang$^2$}

\affiliation{$^1$ Physikalisches Institut (EP3) and
  Institute for Theoretical Physics and Astrophysics,
  University of W$\ddot{u}$rzburg, 97074 W$\ddot{u}$rzburg, Germany;}

\affiliation{$^2$ Department of Physics, McCullough Building, Stanford
  University, Stanford, CA 94305-4045;}

\affiliation{$^2$ Beijing National Laboratory for Condensed Matter
  Physics, and Institute of Physics, Chinese Academy of Sciences,
  Beijing 100190, China;}

\date{\today}

\begin{abstract}
    In this paper we give the full microscopic derivation of the
    model Hamiltonian for the three dimensional topological
    insulators in the $Bi_2Se_3$ family of materials
    ($Bi_2Se_3$, $Bi_2Te_3$ and $Sb_2Te_3$). We first give
    a physical picture to understand the electronic structure by
    analyzing atomic orbitals and applying symmetry principles.
    Subsequently, we give the full microscopic derivation of the model Hamiltonian
    introduced by Zhang {\it et al}~[\onlinecite{zhang2009}] based both on symmetry
    principles and the ${\bf k}\cdot{\bf p}$ perturbation theory.
    Two different types of $k^3$ terms, which break
    the in-plane full rotation symmetry down to three fold rotation
    symmetry, are taken into account. Effective Hamiltonian is derived for the
    topological surface states. Both the bulk and the surface models are investigated
    in the presence of an external magnetic field, and the associated Landau level
    structure is presented. For more quantitative fitting to the
    first principle calculations, we also present a new model Hamiltonian including eight
    energy bands.
\end{abstract}

 \maketitle

\section{Introduction}
\label{sec:intro} Recently, topological insulators (TI) have been
investigated intensively both theoretically and experimentally.
\cite{qi2010,moore2010,hasan2010}
These insulators are fully gapped in the bulk, but have gapless edge or
surface states which are topologically protected by the time
reversal symmetry. Topological insulator was first theoretically
predicted\cite{bernevig2006d} and experimentally
observed\cite{koenig2007} in the $HgTe$ quantum wells. Transport
measurements\cite{koenig2007,roth2009} show the existence of the
gapless edge channel, which demonstrates that $HgTe/CdTe$ quantum
well is a two-dimensional (2D) TI with quantum spin Hall effect.
Later, $Bi_xSb_{1-x}$ was suggested to be a three-dimensional (3D)
TI\cite{fu2007a} with topologically non-trivial surface states,
which were observed by angle-resolved photoemission spectroscopy
(ARPES)\cite{hsieh2008}. However, $Bi_xSb_{1-x}$ has a small
energy gap, alloy disorder and rather complicated surface states.
More recently, new TIs with large bulk gaps $\sim 0.3 eV$ and single
Dirac cone surface states have been theoretically predicted for
$Bi_2Te_3$\cite{zhang2009}, $Sb_2Te_3$\cite{zhang2009} and $Bi_2Se_3$ \cite{zhang2009,xia2009}. ARPES measurement\cite{xia2009,chen2009}
indeed shows the single Dirac cone with linear dispersion around the
$\Gamma$ point in both $Bi_2Se_3$ and $Bi_2Te_3$.
Current research on these materials is developing rapidly.
\cite{zhang2009c,zhang2009e,cheng2010,hanaguri2010,hor2010,hsieh2009b,alpichshev2010,
kong2010,cha2010,peng2010,hsieh2009,roushan2009,chen2010a,butch2010,teweldebrhan2010,
steinberg2010,tang2010,ayala2010,alpichshev2010a}

For deeper understanding and quantitative predictions of novel
phenomena associated with the TIs, it is highly desirable to
construct standard models for both 2D and 3D TIs. Bernevig, Hughes
and Zhang (BHZ)\cite{bernevig2006d} constructed the model
Hamiltonian for the 2D TI in $HgTe$ quantum wells. This model
Hamiltonian demonstrates the basic mechanism of TI behavior through
band inversion induced by spin-orbit coupling (SOC). It has been applied
successfully for quantitative predictions of the helical edge states
and properties under magnetic fields.\cite{koenig2008} 
Zhang {\it et al}\cite{zhang2009} derived a model Hamiltonian for the 3D TI
$Bi_2Se_3$, $Bi_2Te_3$ and $Sb_2Te_3$, and obtained topological
surface states consisting of a single Dirac cone. Interestingly, in
the thin film limit, the 3D TI model reduces exactly to the 2D TI
model by BHZ\cite{liu2010,lu2010,linder2009}. 
In this paper, we give the full microscopic derivation of our model Hamiltonian, first by
constraining its form by symmetry principles and a careful analysis
of the relevant atomic orbitals. Subsequently, we determine the
parameters of our model Hamiltonian by a systematic
${\bf k}\cdot{\bf p}$ expansion near the $\Gamma$ point, and
comparison with the {\it ab initio} calculations\cite{zhang2009}.
Furthermore the higher order $k^3$ terms neglected in Ref.
[\onlinecite{zhang2009}], are also included in the derivation in order to
recover the crystal $C_3$ rotation symmetry\cite{fu2009}. Compared to the symmetry arguments given in Ref.[\onlinecite{zhang2009}], the new derivation
given in this paper determines all the parameters of our model
Hamiltonian by the wavefunctions from {\it ab initio} calculation,
so that no fitting is required and no ambiguity is introduced. As an
application of our model Hamiltonian, we study the bulk and surface
Landau level spectra in a magnetic field. The surface Landau levels
have $\sqrt{B}$ field dependence, as is expected from the Dirac-type
dispersion of the surface states. The gap of 0th Landau level can be
as large as $50meV$ for $10T$ magnetic field, which suggests that
the topological magneto-electric effect\cite{qi2008b,qi2009} can be
observable at such energy scales. Furthermore, we propose a more
quantitative description of the $Bi_2Se_3$ family of TIs by going
beyond the four bands and present a new model Hamiltonian with eight
bands. Recently, our model Hamiltonian has been applied successfully
for understand a number of experiments, including the STM study of
the topological surface states\cite{alpichshev2010,zhang2009c}, STM
study of the surface bound states\cite{alpichshev2010a}, 
STM study of the quasi-particle interference\cite{zhang2009c,alpichshev2010,lee2009},
crossover from 3D to 2D topological insulators\cite{zhang2009e,liu2010,lu2010,linder2009},
and the Landau level of the topological surface states\cite{cheng2010,hanaguri2010}.

The paper is organized as follows. In sec.~\ref{sec:struct} we first
present the lattice structure and the symmetry properties of
Bi$_2$Se$_3$ crystal. Then we turn to the electronic band structure
and discuss about the atomic orbital picture, which is helpful to
capture the essential physics in the long wave length limit. Keeping
such atomic orbital picture in mind, we investigate in detail the
properties of the bands near Fermi surface based on the symmetry
argument in sec.~\ref{sec:Eff}. Furthermore, our model Hamiltonian
for the conduction and valence bands is derived from the theory of
invariants. In sec.~\ref{sec:kp}, we re-derive our model Hamiltonian
from the ${\bf k}\cdot{\bf p}$ theory and determine its parameters
by more fundamental matrix elements of the momentum operator in the
${\bf k}\cdot{\bf p}$ theory. As an application of our model
Hamiltonian, the surface state Hamiltonian and the Landau levels for
both the bulk and surface states are calculated in Sec.~\ref{sec:surface}
and Sec.~\ref{sec:LL} respectively. In Sec.~\ref{sec:8band}, the
quantitative limitation of our model Hamiltonian with four-bands is
discussed and a new model Hamiltonian is proposed to describe the
$Bi_2Se_3$ type of materials more quantitatively. In
Sec.~\ref{sec:conclusion} we provide a brief discussion and
conclusion.

\section{Crystal structure, atomic orbitals and symmetry}
\label{sec:struct}
\begin{figure}[htpb]
   \begin{center}
      \includegraphics[width=3in]{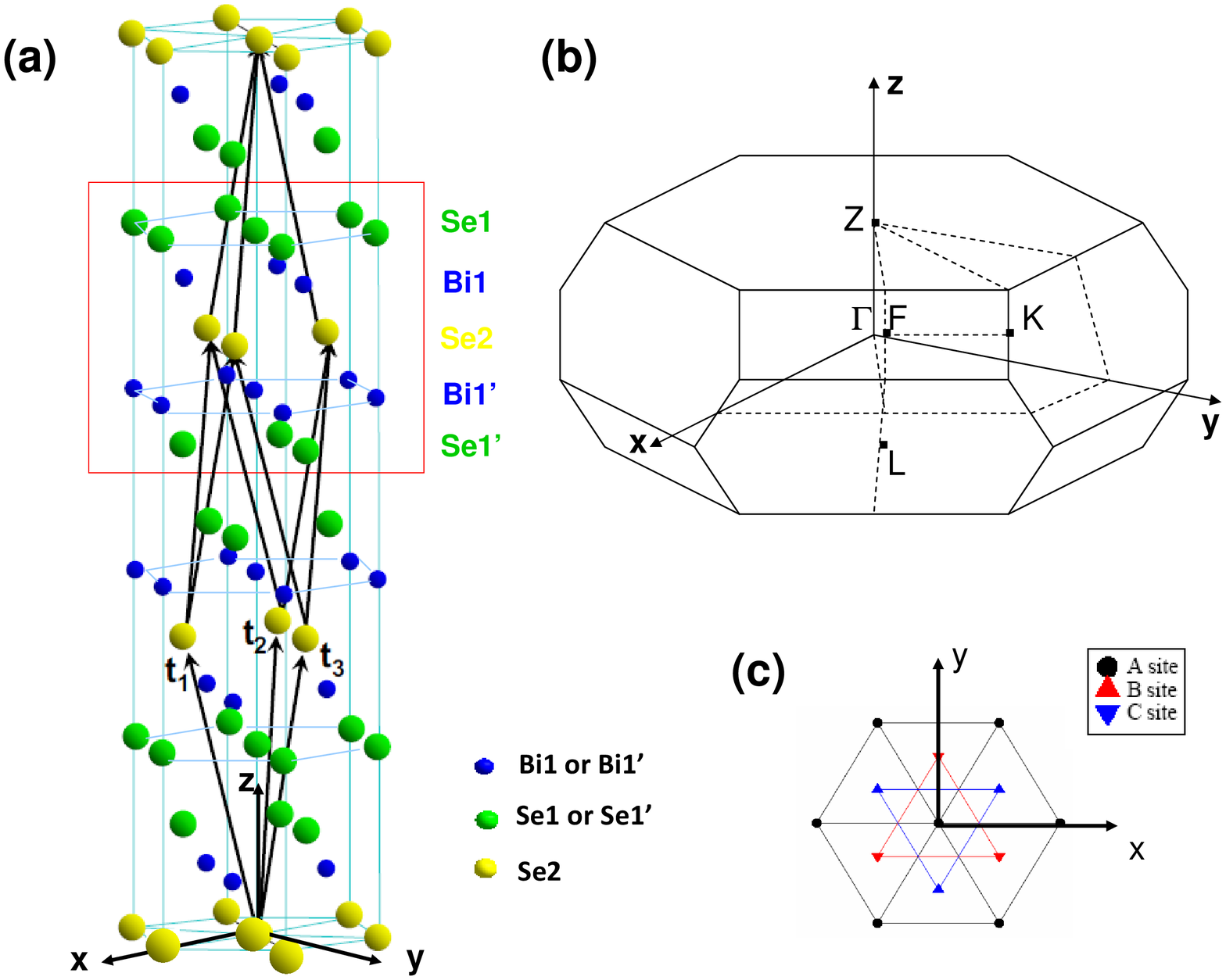}
    \end{center}
    \caption{ (a) The crystal structure of $Bi_2Se_3$. $\vec{t}_{1,2,3}$ is
    the primitive lattice vector, given by $\vec{t}_1=(\sqrt{3}a/3,0,c/3)$,
    $\vec{t}_2=(-\sqrt{3}a/6,a/2,c/3)$,
    $\vec{t}_3=(-\sqrt{3}a/6,-a/2,c/3)$;
    where $a$ is the lattice constant in the x-y plane and $c$ is the lattice
    constant along the z direction.
    The quintuple layer is shown in the red box with $Se1-Bi1-Se2-Bi1'-Se1'$.
    $Se1$ ($Bi1$) and $Se1'$ ($Bi1'$) are equivalent. (b) Brillioun Zone of
    $Bi_2Se_3$. (c) The in-plane triangle lattice has three possible
    position A, B and C. }
    \label{fig:crystal}
\end{figure}

In this section we will describe the crystal structure of $Bi_2Se_3$
family of materials, and discuss the relevant atomic orbitals and
the discrete symmetries. A large portion of the content of this
section is already discussed in Ref. [\onlinecite{zhang2009}], but we feel that
it is helpful to present the more complete version of this
discussion here, to make this paper self-contained. The crystal
structure of $Bi_2Se_3$ is rhombohedral with the space group
$D^5_{3d}$ ($R\bar{3}m$). As shown in Fig. \ref{fig:crystal} (a),
the crystal has layered structure stacked along z-direction with
five atoms (two $Bi$ atoms and three $Se$ atoms) in one unit cell,
including two equivalent $Se$ atoms ($Se1$ and $Se1'$), two equivalent Bi
atoms ($Bi1$ and $Bi1'$) and one $Se$ atom ($Se2$) which is inequivalent to
the $Se1$ and $Se1'$ atoms. Therefore five atomic layers can be viewed
as one unit, which is usually called a quintuple layer. Each atomic
layer forms a triangle lattice, which has three possible positions,
denoted as A, B and C, as shown in Fig.\ref{fig:crystal} (c). Along
the z-direction, the triangle layers are stacked in the order
$A-B-C-A-B-C-\cdots$. We note that the primitive lattice vector
$t_i$ ($i=1,2,3$) is not directed along the z direction. For
example, in one quintuple layer, the $Se2$ atoms occupy the A sites;
in the next quintuple layer, the $Se2$ atoms do not occupy the A sites
but rather the C or B sites. Our coordinate is set as the following:
the origin point is set at the $Se2$ site; z direction is set
perpendicular to the atomic layer, x direction is taken along the
binary axis with the two fold rotation symmetry and y direction is
taken along the bisectrix axis, which is the crossing line of the
reflection plane and the $Se2$ atomic layer plane. The Brillioun
Zone (BZ) of this lattice structure is shown in Fig.
\ref{fig:crystal} (b). This crystal structure has the following
discrete symmetries:
\begin{itemize}
    \item {\it Three fold rotation symmetry $R_3$ along z direction}.
        $R_3$ can be generated by the following transformation:
    $x\rightarrow x\cos\theta-y\sin\theta$,
    $y\rightarrow x\sin\theta+y\cos\theta$ and
    $z\rightarrow z$, where $\theta=\frac{2\pi}{3}$.
\item {\it Two fold rotation symmetry $R_2$ along x direction}.
    $R_2$ corresponds to the following transformation:
        $Se2\rightarrow Se2$, $Bi1\rightarrow Bi1'$,
        $Se1\rightarrow Se1'$; $z\rightarrow -z$,
        $x\rightarrow x$, $y\rightarrow -y$. For this symmetry
    operation, we find that $Bi1$ ($Se1$) and $Bi1'$ ($Se1'$)
    layers inter-change their positions.
\item {\it Inversion $P$}. $P$: $Se2\rightarrow Se2$,
        $Bi1\rightarrow Bi1'$, $Se1\rightarrow Se1'$;
        $z\rightarrow -z$, $x\rightarrow -x$, $y\rightarrow -y$.
    The $Se2$ site is the inversion center
    of this lattice structure, hence we set $Se2$ as the
    origin point. Under inversion operation,
    $Bi1$ ($Se1$) is changed to $Bi1'$ ($Se1'$).
\item {\it Time reversal symmetry $T$}. Time reversal operation is
    given by $T=\Theta K$,
where $\Theta=i\sigma_2$ and $K$ is the complex conjugate operator.
Here $\sigma_{1,2,3}$ are the Pauli matrice for spin.
\end{itemize}

\begin{figure}[htpb]
   \begin{center}
      \includegraphics[width=3.5in]{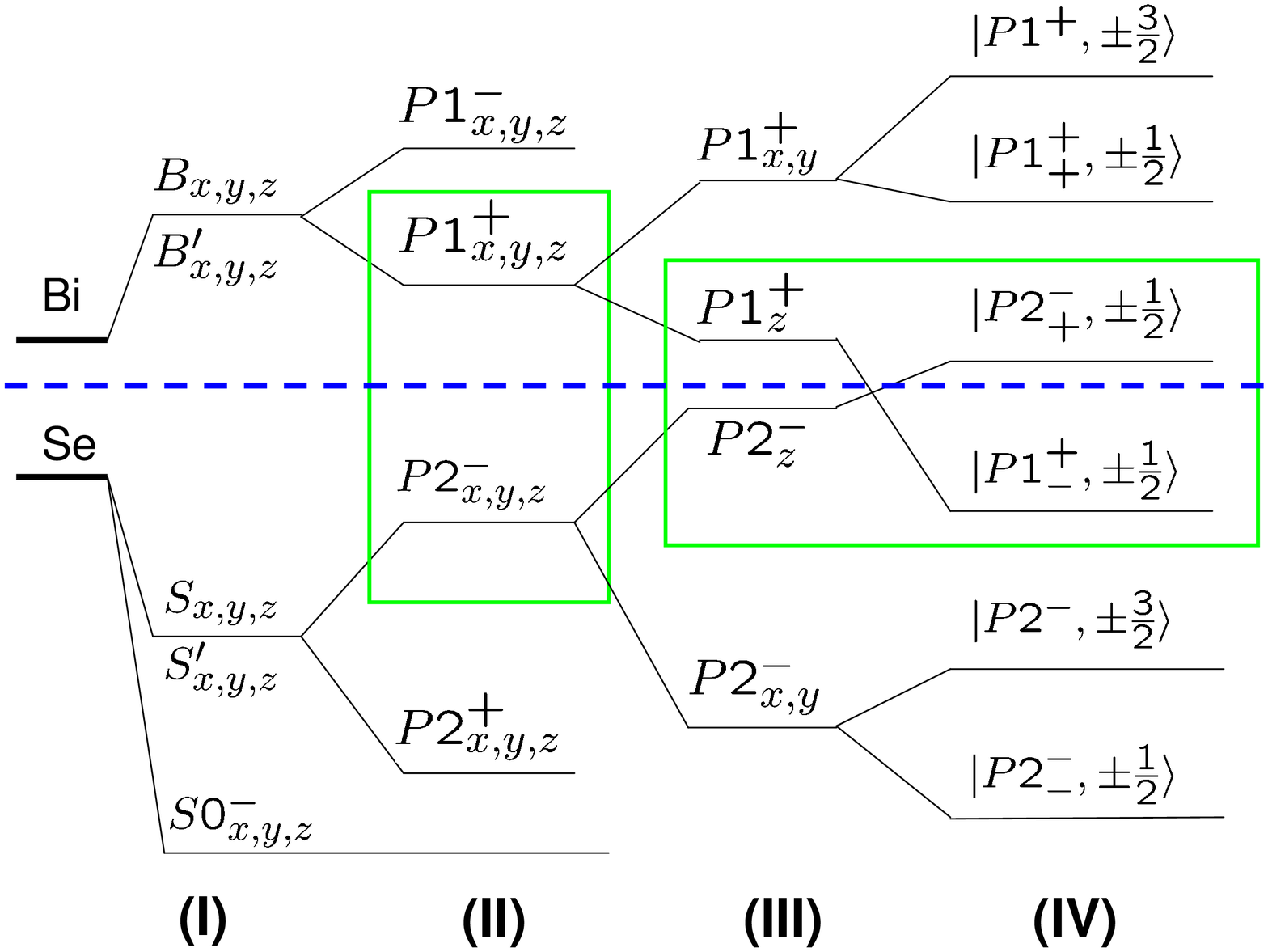}
    \end{center}
    \caption{ Schematic picture of the origin of the band
    structure of $Bi_2Se_3$. Starting from the atomic orbitals
    of $Bi$ and $Se$, the following four steps are required to
    understand the band structure: (I) the hybridization of $Bi$
    orbitals and $Se$ orbitals, (II) the formation of the bonding
    and antibonding states due to the inversion symmetry,
    (III) the crystal field splitting and (IV) the influence of
    the SOC.}
    \label{fig:atomic}
\end{figure}

In order to get a physical picture of the band structure of
$Bi_2Se_3$, we start from the atomic orbitals of $Bi$ and $Se$. The
electron configuration of $Bi$ is $6s^26p^3$ and that of $Se$ is
$4s^24p^4$. The outmost shells for both $Bi$ and $Se$ are p-orbital,
therefore it is natural to consider only p-orbitals of $Bi$ and $Se$
and neglect other orbitals. As discussed above, $Bi_2Se_3$ has a
layered structure. The chemical bonding is very strong within one
quintuple layer but the two neighboring quintuple layers are only
coupled by the van der Waals force. Therefore it is reasonable for
us to first focus on one quintuple layer. Within one quintuple layer
there are 5 atoms in one unit cell and each atom has three orbitals
($p_x$, $p_y$ and $p_z$), therefore totally there are 15 orbitals.
The spin is neglected first and will be discussed later when we
introduce SOC into the system. We denote these
orbitals as $|\Lambda,\alpha\rangle$ with $\Lambda=Bi1$, $Bi1'$,
$Se1$, $Se2$, $Se1'$ and $\alpha=p_x$, $p_y$, $p_z$. As shown in
Fig.\ref{fig:crystal} (a), the $Se2$ atomic layer stays in the
middle of the quintuple layer and is sandwiched by two $Bi$ layers
($Bi1$ and $Bi1'$), while two $Se$ layers ($Se1$ and $Se1'$) are
located at the outermost. Since all the $Se$ layers are seperated by
$Bi$ layers, the strongest coupling in this system is the coupling
between $Bi$ layers and $Se$ layers. Such coupling causes level
repulsion, so that the $Bi$ energy levels are pushed up and form
new hybridized states $|B_\alpha\rangle$ and $|B'_\alpha\rangle$
while the $Se$ energy levels are pushed down and yield three
states $|S_\alpha\rangle$, $|S'_\alpha\rangle$ and
$|S0_\alpha\rangle$, as shown in Fig.\ref{fig:atomic} (I). Since the
system has inversion symmetry, it is convenient to combine these
orbitals to form the bonding and anti-bonding states with the
definite parity, which are given by
\begin{eqnarray}
    |P1^\pm,\alpha\rangle=\frac{1}{\sqrt{2}}(|B_\alpha\rangle\mp|B'_\alpha\rangle)\label{eq:struct_P1P21}\nonumber\\
        |P2^\pm,\alpha\rangle=\frac{1}{\sqrt{2}}(|S_\alpha\rangle\mp|S'_\alpha\rangle)\label{eq:struct_P1P22}
\end{eqnarray}
with the upper index denoting the parity and $\alpha=p_x,p_y,p_z$.
When the coupling between $|B_\alpha(S_\alpha)\rangle$ and
$|B'_\alpha(S'_\alpha)\rangle$ is taken into account, the bonding
and anti-bonding states are split, with the anti-bonding state
having higher energy than the bonding state. Therefore as shown in
Fig.\ref{fig:atomic} (II), the states $|P1^+,\alpha\rangle$ and
$|P2^-,\alpha\rangle$ are found to be near the Fermi surface, hence
we focus on $|P1^+,\alpha\rangle$ and $|P2^-,\alpha\rangle$
($\alpha=p_x,p_y,p_z$) and neglect the other states. Furthermore the
crystal has layered structure, so the z direction is different from
the x or y directions in the atomic plane. Thus there is an energy
splitting between $p_z$ and $p_{x,y}$ orbitals for both $P1^+$ and
$P2^-$ states. We find that $|P1^+,p_{x,y}\rangle$ orbitals have
higher energy than $|P1^+,p_z\rangle$, while $|P2^-,p_{x,y}\rangle$
orbitals have lower energy than $|P2^-,p_z\rangle$. Consequently,
the conduction band mainly consists of $|P1^+,p_z\rangle$ while the
valence band is dominated by the $|P2^-,p_z\rangle$ orbital before
SOC is considered, as shown in Fig.\ref{fig:atomic} (III).

Next we include SOC effect in the above atomic picture. The
states $|P1^+,\alpha,\sigma\rangle$ and $|P2^-,\alpha,\sigma\rangle$
are all double degenerate, with one more index
$\sigma=\uparrow,\downarrow$ to denote spin. The atomic SOC
Hamiltonian is given by $\hat{H}_{so}=\lambda{\bf s}\cdot {\bf L}$
with $\lambda=\frac{1}{2m_0^2c^2}\frac{1}{r}\frac{\partial U}{\partial r}$
depending on the detail potential $U$ of atoms,
which couples orbital angular momentum to spin. It is
convenient to transform the $p_x$ and $p_y$ orbitals to $p_\pm$ with
definite orbital angular momentum:
\begin{eqnarray}
    &&|\Lambda,p_+,\sigma\rangle=-\frac{1}{\sqrt{2}}(|\Lambda,p_x,
    \sigma\rangle+i|\Lambda,p_y,\sigma\rangle),\\
        &&|\Lambda,p_-,\sigma\rangle=\frac{1}{\sqrt{2}}(|\Lambda,p_x,
    \sigma\rangle-i|\Lambda,p_y,\sigma\rangle),
    \label{eq:struct_porbital}
\end{eqnarray}
where $\Lambda=P1^+,P2^-$. Within this basis, the atomic SOC
Hamiltonian is given by
\begin{eqnarray}
    &&\langle\Lambda,p_+,\uparrow|H_{so}|\Lambda,p_+,\uparrow\rangle
=\langle\Lambda,p_-,\downarrow|H_{so}|\Lambda,p_-,\downarrow\rangle
\equiv\frac{\lambda_\Lambda}{2}\nonumber\\
 &&\langle\Lambda,p_+,\downarrow|H_{so}|\Lambda,p_+,\downarrow\rangle
=\langle\Lambda,p_-,\uparrow|H_{so}|\Lambda,p_-,\uparrow\rangle
\equiv-\frac{\lambda_\Lambda}{2}\nonumber\\
  &&\langle\Lambda,p_+,\downarrow|H_{so}|\Lambda,p_z,\uparrow\rangle
    =\langle\Lambda,p_-,\uparrow|H_{so}|\Lambda,p_z,\downarrow\rangle
    \equiv\frac{\lambda_\Lambda}{\sqrt{2}}\nonumber\\
 &&\langle\Lambda,p_z,\uparrow(\downarrow)|H_{so}|\Lambda,p_z,
    \uparrow(\downarrow)\rangle=0.
    \label{eq:struct_soc1}
\end{eqnarray}
Here the value of $\lambda_\Lambda$ is a linear combination of
the SOC coefficient for $Bi$ and $Se$, depending
on how much the orbitals of $Bi$ and $Se$ are mixed into the state
$|\Lambda\rangle$. The sign of $\lambda_\Lambda$ is always positive
for $\Lambda=P1^+, P2^-$ since the potential is always attractive for atoms.
As we see, since the total angular momentum along the z direction is still
conserved, hybridization only occurs between
$|\Lambda,p_z,\uparrow\rangle$ ($|\Lambda,p_z,\downarrow\rangle$)
and $|\Lambda,p_+,\downarrow\rangle$
($|\Lambda,p_-,\uparrow\rangle$). After taking into account SOC, the
new eigen-states are given by
\begin{eqnarray}
    &&|\Lambda,\frac{3}{2}\rangle=|\Lambda,p_+,\uparrow\rangle\label{eq:struct_SOCwf1}\\
    &&|\Lambda,-\frac{3}{2}\rangle=|\Lambda,p_-,\downarrow\rangle\label{eq:struct_SOCwf2}\\
    &&|\Lambda_+,\frac{1}{2}\rangle=u^{\Lambda}_{+}|\Lambda,p_z,\uparrow\rangle
    +v^{\Lambda}_{+}|\Lambda,p_+,\downarrow\rangle\label{eq:struct_SOCwf3}\\
    &&|\Lambda_-,\frac{1}{2}\rangle=u^{\Lambda}_{-}|\Lambda,p_z,\uparrow\rangle
    +v_{-}^{\Lambda}|\Lambda,p_+,\downarrow\rangle\label{eq:struct_SOCwf4}\\
    &&|\Lambda_+,-\frac{1}{2}\rangle=(u^\Lambda_+)^*|\Lambda,p_z,\downarrow\rangle
    +(v^\Lambda_+)^*|\Lambda,p_-,\uparrow\rangle\label{eq:struct_SOCwf5}\\
    &&|\Lambda_-,-\frac{1}{2}\rangle=(u^\Lambda_-)^*|\Lambda,p_z,\downarrow\rangle
    +(v^\Lambda_-)^*|\Lambda,p_-,\uparrow\rangle\label{eq:struct_SOCwf6}
\end{eqnarray}
with the eigen-energies $E^{\Lambda}_{3/2}$ and $E^{\Lambda_\pm}_{1/2}$
(each is double degenerate) and $u$, $v$ obtained by solving the following
$2\times2$ Hamiltonian
\begin{eqnarray}
    \hat{H}=\left(
    \begin{array}{cc}
        E_{\Lambda,x}-\lambda_\Lambda/2&\lambda_\Lambda/\sqrt{2}\\
        \lambda_\Lambda/\sqrt{2}&E_{\Lambda,z}
    \end{array}
    \right).
    \label{eq:struct_Hso1}
\end{eqnarray}
For the above eigen-states (\ref{eq:struct_SOCwf1})$\sim$(\ref{eq:struct_SOCwf6}),
all the information about SOC is included in the coefficient $u$ and $v$, which
are given by
\begin{eqnarray}
    \left(
    \begin{array}{c}
        u^\Lambda_\pm\\
        v^\Lambda_\pm
    \end{array}
    \right)=\frac{1}{N_\pm}\left(
    \begin{array}{c}
        \Delta E_{\Lambda}\pm\sqrt{\left( \Delta E_{\Lambda}  \right)^2
        +\frac{\lambda_\Lambda^2}{2}}\\
        \lambda_\Lambda/\sqrt{2}
    \end{array}
    \right)
    \label{eq:struct_uv}
\end{eqnarray}
explicitly, where $N_\pm=\lambda_\Lambda^2+2\Delta E_{\Lambda}^2\pm2\Delta E_{\Lambda}
\sqrt{\Delta E_{\Lambda}^2+\lambda_\Lambda^2/2}$ and $\Delta E_{\Lambda}
=\frac{E_{\Lambda,x}-E_{\Lambda,z}-\lambda_\Lambda/2}{2}$. The energy splitting
between the $p_{x(y)}$ orbital and the $p_z$ orbital due to the crystal field
is larger than the energy scale of SOC and $\Delta E_\Lambda$ is dominated by
$E_{\Lambda,x}-E_{\Lambda,z}$.
Now as we see, the SOC couples $|\Lambda,p_z,\uparrow\rangle$
($|\Lambda,p_z,\downarrow\rangle$) to
$|\Lambda,p_+,\downarrow\rangle$ ($|\Lambda,p_-,\uparrow\rangle$) so
that it induces the level repulsion between these two states.
Consequently, $|P1^+_-,\pm\frac{1}{2}\rangle$ is pushed down while
$|P2^-_+,\pm\frac{1}{2}\rangle$ is pushed up, which yields the level
crossing between these two pairs of states, when the SOC is strong
enough, as shown in Fig.\ref{fig:atomic} (IV). Since these two pairs
of states have the opposite parity, their crossing leads to a {\it
band inversion}, similar to the case in the $HgTe$ quantum
wells\cite{bernevig2006d}. This is the key signature of the
topological insulator phase in $Bi_2Se_3$ family of
materials\cite{zhang2009}. Therefore in the following we will focus
on these four states and regard the other states as the
perturbation.
%

\section{Model Hamiltonian derived from symmetry principles}
\label{sec:Eff}
\begin{figure}[htpb]
   \begin{center}
      \includegraphics[width=3.0in]{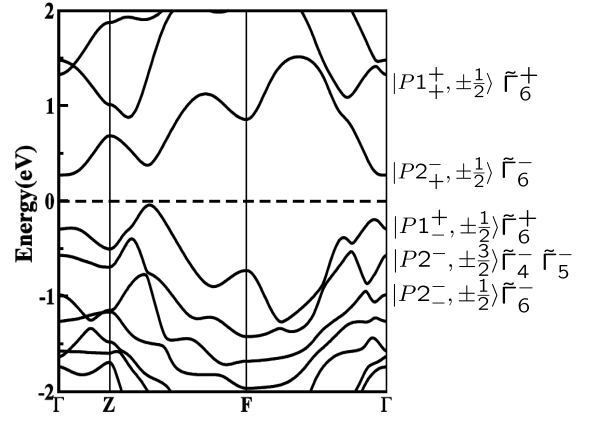}
    \end{center}
    \caption{ The band structure of $Bi_2Se_3$ is obtained
    from {\it ab initio} calculation, and the bands near Fermi
    surface is identified with $|\Lambda^\pm,\alpha\rangle$.
    $\Lambda=P1_\pm,P2_\pm$ and $\alpha=\pm\frac{1}{2},
    \pm\frac{3}{2}$. The corresponding irreducible representation
    is also given. }
    \label{fig:band}
\end{figure}

From the discussion of the atomic orbitals in the last section, we
obtain an intuitive physical picture of the band structure of
$Bi_2Se_3$. Compared with the {\it ab initio} calculation, we can
denote the bands near Fermi surface by $|\Lambda^\pm,\alpha\rangle$
where $\Lambda=P1_\pm,P2_\pm$ and
$\alpha=\pm\frac{1}{2},\pm\frac{3}{2}$, as shown in Fig
\ref{fig:band}. Roughly, these states mainly consist of the bonding
or anti-bonding states of the p-orbitals of $Bi$ or $Se$ atoms.
However, other orbitals such as s-orbitals of $Bi$ and $Se$ will
also mix into these states. To identify each band without any
ambiguity, it is necessary to relate each band with the
representation of the crystal symmetry. At $\Gamma$ point, each
state should belong to an irreducible representation of the crystal
symmetry group and the hybridization between orbitals preserve the
symmetry properties. Therefore, a suitable method to identify each
band is to use the symmetry of the crystal. In this section, we will
first identify each band according to the irreducible representation
of the crystal group $D^5_{3d}$ and then try to derive our model
Hamiltonian just from symmetry principles.

First let's consider the states without spin, which are denoted as
$|\Lambda^\pm,\alpha\rangle$ with $\Lambda=P1,P2$ and
$\alpha=p_x,p_y,p_z$. The crystal of $Bi_2Se_3$ belongs to the group
$D^5_{3d}$ with the character table given in table
(\ref{tab:Character1}) of Appendix
\ref{app:A_symmetry}\cite{dresselhausbook2008}. Since the crystal is
inversion symmetric, each representation has a definite parity
eigenvalue. For each parity, there are two one-dimensional
representations $\tilde\Gamma^{\pm}_1$ and $\tilde\Gamma^{\pm}_2$
and one two-dimensional representation $\tilde\Gamma^\pm_3$, where
the upper index denotes the parity (+ for even and - for odd).
According to the wave functions constructed from the simple atomic
orbital picture, we can determine the transformation property of the
wave functions under the generators $R_3$, $R_2$ and $P$ of the
point group. For example, let's look at the operation $R_2$ on the
state
$|P1^+,p_x\rangle=\frac{1}{\sqrt{2}}(|B_x\rangle-|B'_x\rangle)$. The
$R_2$ rotation does not change the $p_x$ orbital, however it changes
the position of $Bi1$ ($Se1$) and $Bi1'$ ($Se1'$) and
correspondingly changes $|B\rangle$ to $|B'\rangle$, thus we should
have $R_2|P1^+,p_x\rangle=-|P1^+,p_x\rangle$. Similiar argument can
be applied to other states and finally the transformation of the
states under the cyrstal symmetry operation is listed as follows.
\begin{itemize}
    \item {\it Three fold rotation symmetry} $R_3$: $|\Lambda^{\pm},p_x\rangle
         \rightarrow \cos\theta|\Lambda^{\pm},p_x\rangle-\sin\theta
         |\Lambda^{\pm},p_y\rangle$, $|\Lambda^{\pm},p_y\rangle\rightarrow
         \sin\theta|\Lambda^\pm,p_x\rangle+\cos\theta|\Lambda^\pm,p_y\rangle$
         and $|\Lambda^\pm,p_z\rangle\rightarrow|\Lambda^\pm,p_z\rangle$,
         with $\theta=\frac{2\pi}{3}$.
     \item {\it Two fold rotation symmetry} $R_2$: $|\Lambda^{\pm},p_x\rangle\rightarrow
        \mp|\Lambda^{\pm},p_x\rangle$, $|\Lambda^{\pm},p_y\rangle\rightarrow
        \pm|\Lambda^{\pm},p_y\rangle$, $|\Lambda^{\pm},p_z\rangle\rightarrow
        \pm|\Lambda^{\pm},p_z\rangle$.
    \item {\it Inversion }$P$: $|\Lambda^{\pm},\alpha\rangle\rightarrow
        \pm|\Lambda^{\pm},\alpha\rangle$, $\alpha=p_x,p_y,p_z$.
\end{itemize}
Here $\Lambda=P1_\pm,P2_\pm$. According to the above transformation,
we find that $|\Lambda^{+(-)},p_x\rangle$ and
$|\Lambda^{+(-)},p_y\rangle$ belong to the
$\tilde\Gamma^{+(-)}_3$representation. $|\Lambda^{+},p_z\rangle$
belongs to $\tilde\Gamma_1^{+}$ representation and
$|\Lambda^{-},p_z\rangle$ belongs to $\tilde\Gamma_2^-$
representation.


To take into account spin, we introduce the spinor
representation $\tilde\Gamma^+_6$, which changes its sign under the
rotation $\mathcal{C}=2\pi$. The double group of $D^5_{3d}$ can be
constructed by the direct product of $\tilde\Gamma_{1,2,3}^\pm$ and
$\tilde\Gamma_6^+$. As shown in (\ref{eq:App_Dirpro1}) $\sim$
(\ref{eq:App_Dirpro3}), we find that
$\tilde\Gamma_3^\pm\otimes\tilde\Gamma_6^+$ will give two new
one-dimensional representations $\tilde\Gamma^\pm_4$ and
$\tilde\Gamma^\pm_5$, which are conjugate to each other. The
character table of the double group for $D^5_{3d}$ is given in table
(\ref{tab:Character2}) of Appendix
\ref{app:A_symmetry}\cite{dresselhausbook2008}. With SOC, the
eigen-states in (\ref{eq:struct_SOCwf1}) $\sim$
(\ref{eq:struct_SOCwf2}) can also be analysed by the decomposition
of direct production. From (\ref{eq:App_Dirpro2}) and
(\ref{eq:App_Dirpro3}), the direct product of $\tilde\Gamma^\pm_6$
and $\tilde\Gamma^\pm_{1,2}$ always gives $\tilde\Gamma^\pm_6$
representation, therefore $|\Lambda^+,\pm\frac{1}{2}\rangle$ with
$\Lambda=P1,P2$ should belong to $\tilde\Gamma^+_6$ representaion
while $|\Lambda^-,\pm\frac{1}{2}\rangle$ should belong to
$\tilde\Gamma^-_6$ representation. The states
$|\Lambda^\pm,\pm3/2\rangle$ originate from the combination of
$|\Lambda,p_{x,y}\rangle$ and spin. According to
(\ref{eq:App_Dirpro1}), it is expected that
$|\Lambda^\pm,\pm3/2\rangle$ should be a combination of
$\tilde\Gamma_4^\pm$ and $\tilde\Gamma_5^\pm$ representations.
Indeed by carefully inspecting the transformation behavior under the
operation $R_2$ and $R_3$, we find that
\begin{eqnarray}
    |\Lambda^\pm,\tilde\Gamma_4\rangle=\frac{1}{\sqrt{2}}
    (|\Lambda^\pm,3/2\rangle+|\Lambda^\pm,-3/2\rangle)
    \label{eq:eff_Gamma4}
\end{eqnarray}
belongs to the $\tilde\Gamma_4^\pm$ representation,
while
\begin{eqnarray}
    |\Lambda^\pm,\tilde\Gamma_5\rangle=\frac{1}{\sqrt{2}}
    (|\Lambda^\pm,3/2\rangle-|\Lambda^\pm,-3/2\rangle)
    \label{eq:eff_Gamma5}
\end{eqnarray}
belongs to the $\tilde\Gamma_5^\pm$ representation.
The above results can also be worked out by considering the forms
of the transformation for the states (\ref{eq:struct_SOCwf1}) $\sim$
(\ref{eq:struct_SOCwf2}), which are given by

\begin{itemize}
    \item {\it Three fold rotation symmetry} $R_3$: $|\Lambda,\pm\frac{1}{2}\rangle
        \rightarrow e^{\pm i\frac{\pi}{3}}|\Lambda,\pm\frac{1}{2}\rangle$,
        $|\Lambda,\pm\frac{3}{2}\rangle\rightarrow-|\Lambda,\pm\frac{3}{2}\rangle$,
        where $\Lambda=P1^\pm_\pm,P2^\pm_\pm$.
    \item {\it Two fold rotation symmetry} $R_2$: $|\Lambda^+,\pm\frac{1}{2}\rangle
        \rightarrow i|\Lambda^+,\mp\frac{1}{2}\rangle$, $|\Lambda^-,\pm\frac{1}{2}\rangle
        \rightarrow -i|\Lambda^-,\mp\frac{1}{2}\rangle$, $|\Lambda^+,\pm\frac{3}{2}\rangle
        \rightarrow i|\Lambda^+,\mp\frac{3}{2}\rangle$, $|\Lambda^-,\pm\frac{3}{2}\rangle
        \rightarrow -i|\Lambda^-,\mp\frac{3}{2}\rangle$, with $\Lambda=P1_\pm,P2_\pm$.
    \item {\it Inversion }$P$: $|\Lambda^{\pm},\alpha\rangle\rightarrow
        \pm|\Lambda^{\pm},\alpha\rangle$, with $\Lambda=P1_\pm,P2_\pm$
        and $\alpha=\pm\frac{3}{2},\pm\frac{1}{2}$.
\end{itemize}

It is instructive to compare the present case with the more common
semiconductor crystal structures, such as diamond or zinc-blende
structure. In that case, the coupling between p orbitals and the
spin usually gives the four-dimensional $\tilde\Gamma_8$ and the
two-dimensional $\tilde\Gamma_7$ representation. In the present
case, due to the lower symmetry of the crystal structure, the
$\tilde\Gamma_7$ representation is the same as the $\tilde\Gamma_6$
representation while the $\tilde\Gamma_8$ representation is reduced
to two one-dimensional representations $\tilde\Gamma_4$ and
$\tilde\Gamma_5$ and one two-dimensional representation
$\tilde\Gamma_6$. In Fig \ref{fig:band}, the representation of the
bands near the Fermi surface is given.

Next we derive our model Hamiltonian to describe the low energy
physics of $Bi_2Se_3$ just based on the symmetry of the wave
function at $\Gamma$ point. As described above, near the Fermi
surface the conduction band and valence band are determined by the
four states $|P1^+_-,\pm\frac{1}{2}\rangle$ and
$|P2^-_+,\pm\frac{1}{2}\rangle$, belonging to the $\tilde\Gamma_6^+$
and $\tilde\Gamma_6^-$ representation. Therefore the minimum model
Hamiltonian for $Bi_2Se_3$ should be written with these four states
as the basis. Generally, any $4\times 4$ Hamiltonian can be expanded
with Dirac $\Gamma$ matrices as
\begin{eqnarray}
    \hat{H}_{eff}=\epsilon({\bf k}){\bf I}+\sum_id_i({\bf k})\Gamma_i
    +\sum_{ij}d_{ij}({\bf k})\Gamma_{ij}
    \label{eq:Eff_Heff}
\end{eqnarray}
where ${\bf I}$ is the $4\times4$ identity matrix, $\Gamma_i$
($i=1\cdots 5$) denote the five Dirac $\Gamma$ matrices satisfying
$\left\{ \Gamma_i,\Gamma_j \right\}=2\delta_{ij}$, and the ten
anti-commutators of $\Gamma$ matrices are given by
$\Gamma_{ij}=[\Gamma_i,\Gamma_j]/2i$. $\epsilon({\bf k})$, $d_i({\bf
k})$ and $d_{ij}({\bf k})$ can be expanded by the powers of the
momentum ${\bf k}$. The construction of $\Gamma$ is given in
Appendix \ref{app:B_Gamma}. Now let's assume the above Hamiltonian
is written in the basis $|P1^+_-,\frac{1}{2}\rangle$,
$|P2^-_+,\frac{1}{2}\rangle$, $|P1^+_-,-\frac{1}{2}\rangle$ and
$|P2^-_+,-\frac{1}{2}\rangle$. Then according to the transformation
of the states under the symmetry operation discussed above, we can
construct the transformation matrices as follows.
\begin{itemize}
\item {\it Time reversal symmetry:} $T=\Theta K$,
where $\Theta=i\sigma_2\otimes1$ and $K$ is the complex conjugate operator.
\item {\it Three fold rotation symmetry operation:}
    $R_3=e^{i\frac{\Pi}{2}\theta}$
    with $\Pi=\sigma_3\otimes 1$ and $\theta=\frac{2\pi}{3}$.
\item {\it Two fold rotation symmetry operation:}
    $R_2=i\sigma_1\otimes\tau_3$.
\item {\it Inversion:} $P=1\otimes \tau_3$.
\end{itemize}
In the above, $\sigma$ acts in the spin basis and $\tau$ acts in the
basis of $P1^+$ and $P2^-$ sub-bands. According to the above
transformation matrices, we can obtain the irreducible
representation of each $\Gamma$ matrix, details of which are derived
in Appendix \ref{app:B_Gamma}. The invariance of the Hamiltonian
requires that the function $d_i({\bf k})$ ($d_{ij}({\bf k})$) should
have the same behavior to the corresponding $\Gamma_i$
($\Gamma_{ij}$) matrix under the symmetry operation, which means
that they should belong to the same representation of the crystal
point group. In the table \ref{tab:Character3} of the Appendix
\ref{app:B_Gamma}, we list the representation for both the $\Gamma$
matrices and the polynomials of ${\bf k}$, and also their
transformation properties under the time reversal operation. Since
we hope to preserve both time reversal symmetry and crystal
symmetry, we must choose the $\Gamma$ matrices and polynomials of
${\bf k}$ with the same representation. For example, $\Gamma_1$ and
$\Gamma_2$ carry the representation $\tilde{\Gamma}^-_3$ and are odd
under time reversal, and so are $k_x$ and $k_y$. Therefore they can
together form an invariant term for Hamiltonian. Finally, up to
$O(k^3)$, our model Hamiltonian yields
\begin{eqnarray}
    &&H'_{eff}=H'_0+H'_3\nonumber\\
    &&H'_0=\epsilon_{{\bf k}}+\mathcal{M}({\bf k})\Gamma_5
    +\mathcal{B}(k_z)\Gamma_4k_z\nonumber\\
    &&+\mathcal{A}(k_\parallel)(\Gamma_1k_y-\Gamma_2k_x)\\
    &&H'_3=R_1\Gamma_3(k_x^3-3k_xk_y^2)+R_2\Gamma_4(3k_x^2k_y-k_y^3)
    \label{eq:Ham}
\end{eqnarray}
where $\epsilon_{{\bf k}}=C_0+C_1k_z^2+C_2k^2_\parallel$,
and $\mathcal{M}({\bf k})=M_0+M_1k_z^2+M_2k^2_\parallel$ and
$\mathcal{A}(k_\parallel)=A_0+A_2k^2_\parallel$ and
$\mathcal{B}(k_z)=B_0+B_2k^2_z$ and $k^2_\parallel=k_x^2+k_y^2$.
$H'_0$ keeps the in-plane rotation symmetry along z direction, while $H'_3$ breaks the
in-plane rotation symmetry down to three fold rotation symmetry. In
the following, we find the bulk $H'_3$ term will also lead to the
correction to the effective surface Hamiltonian, which has been studied in Ref. [\onlinecite{fu2009}].

The above Hamiltonian is the same to that presented by Zhang {\it et
al} in the Ref. [\onlinecite{zhang2009}], which can be shown by performing the
transformation
\begin{eqnarray}
    U_1=\left(
    \begin{array}{cccc}
        1&0&0&0\\
        0&-i&0&0\\
        0&0&1&0\\
        0&0&0&i
    \end{array}
    \right),
    \label{eq:tranU1}
\end{eqnarray}
and the Hamiltonian is transformed into
\begin{eqnarray}
    &&H_{eff}=H_0+H_3\nonumber\\
    &&H_0=U_1H'_0U_1^\dag=\epsilon_{{\bf k}}+\nonumber\\
    &&\left(
    \begin{array}{cccc}
        \mathcal{M}({\bf k})&\mathcal{B}(k_z)k_z&0&\mathcal{A}(k_{\parallel})k_-\\
        \mathcal{B}(k_z)k_z&-\mathcal{M}({\bf k}) &\mathcal{A}(k_{\parallel})k_-&0\\
        0&\mathcal{A}(k_{\parallel})k_+&\mathcal{M}({\bf k})&-\mathcal{B}(k_z)k_z\\
        \mathcal{A}(k_{\parallel})k_+&0&-\mathcal{B}(k_z)k_z&-\mathcal{M}({\bf k})
    \end{array}
    \right)\label{eq:Ham0}\\
    &&H_3=U_1H'_3U_1^\dag=\frac{R_1(k_+^3+k_-^3)}{2}\left(
    \begin{array}{cccc}
        0&i&0&0\\
        -i&0&0&0\\
        0&0&0&i\\
        0&0&-i&0
    \end{array}
    \right)\nonumber\\
    &&+\frac{R_2(k_+^3-k_-^3)}{2}\left(
    \begin{array}{cccc}
        0&-i&0&0\\
        -i&0&0&0\\
        0&0&0&i\\
        0&0&i&0
    \end{array}
    \right)
    \label{eq:Ham3}
\end{eqnarray}
Now we can see that $H_0$ is nearly the same to the model
Hamiltonian (1) in Ref. [\onlinecite{zhang2009}], except the $A_2$ term and
$B_2$ term, which represent the high order correction to the Fermi
velocity $A_0$ and $B_0$. Since such correction is not important
near $\Gamma$ point, we will neglect these two terms in the
following. Derivation of our model Hamiltonian (\ref{eq:Ham0}) and
(\ref{eq:Ham3}) from the symmetry principles is the central result
of this section.

\section{Model Hamiltonian derived from the ${\bf k}\cdot{\bf p}$
perturbation theory} \label{sec:kp}
Up to now we have obtained our model Hamiltonian from the symmetry
principles, or the theory of invariants\cite{winklerbook2003}. In
this section, we will derive the model Hamiltonian through another
way, ${\bf k}\cdot{\bf p}$ theory, and connect the parameters of
the model Hamiltonian to the more fundamental matrix elements of
momentum in ${\bf k}\cdot{\bf p}$ theory.

The basic idea of ${\bf k}\cdot{\bf p}$ theory is to use the wave
function at $\Gamma$ point in BZ as the zeroth-order wave function
and treat
\begin{eqnarray}
    \hat{H}'=\frac{\hbar}{m_0}{\bf k}\cdot{\bf p}
    \label{eq:kp_H1}
\end{eqnarray}
as a perturbation, where ${\bf p}=-i\hbar\partial_{{\bf r}}$ is
the momentum operator acting on the zeroth-order wave function and
the crystal momentum ${\bf k}$ is regarded as a small parameter
during the perturbation procedure. The model Hamiltonian is expanded
by the powers of $k$. With the perturbation formalism
(\ref{eq:EH1_pert0}) $\sim$ (\ref{eq:EH1_pert3}), we can project the
system into the subspace spanned by the four states
$|P1^+_-,1/2\rangle\equiv|1\rangle$,
$|P2^-_+,1/2\rangle\equiv|2\rangle$,
$|P1^+_-,-1/2\rangle\equiv|3\rangle$ and
$|P2^-_+,-1/2\rangle\equiv|4\rangle$, which are used as the basis of
our model Hamiltonian. All the other states are treated in the
perturbation procedure and the details are given in Appendix
\ref{app:C_kp}. The obtained model Hamiltonian will depend on a
series of matrix elements of momentum $\langle
\Lambda_1,\alpha|{\bf p}|\Lambda_2, \beta\rangle$, which can be
simplified due the symmetry of the crystal. For example, due to the
existence of the inversion symmetry, all the states at $\Gamma$ point
have definite parity eigenvalues. Since the momentum ${\bf p}$ has odd parity,
the matrix elements of momentum between two states with the same
parity always vanish. The wave function at the $\Gamma$ point can be
obtained through {\it ab inito} calculation, consequently all these
matrix elements can be calculated. With these matrix elements, we
apply the perturbation formulism (\ref{eq:EH1_pert0}) $\sim$
(\ref{eq:EH1_pert3}) to the system and recover our model Hamiltonian
(\ref{eq:Ham0}) and (\ref{eq:Ham3}). The parameters of our model
Hamiltonian $C_0$, $C_1$, $C_2$, $M_0$, $M_1$, $M_2$, $A_0$, $B_0$,
$R_1$ and $R_2$ can be expressed as the function of the parameters
$P_{\Lambda_1,\Lambda_2}$, $Q_{\Lambda_1,\Lambda_2}$,
$M_{\Lambda_1,\Lambda_2}$, $N_{\Lambda_1,\Lambda_2}$,
$R_{\Lambda_1,\Lambda_2}$ and $S_{\Lambda_1,\Lambda_2}$ through
(\ref{eq:appC_para1}) $\sim$ (\ref{eq:appC_para10}) in the Appendix
\ref{app:C_kp}. With these expressions, we can numerically calculate
the values of the parameters of our model Hamiltonian, which is
listed in table \ref{tab:4b_parameter}. We note that the parameters
given here are different from those in Ref. [\onlinecite{zhang2009}]
and [\onlinecite{shan2010}], where the parameters are determined
by fitting to the energy dispersion, which has some ambiguity.
In the present method, since we directly calculate the matrix elements
of momentum from microscopic wave functions, there is no ambiguity.

\begin{figure}[htpb]
   \begin{center}
      \includegraphics[width=3.5in]{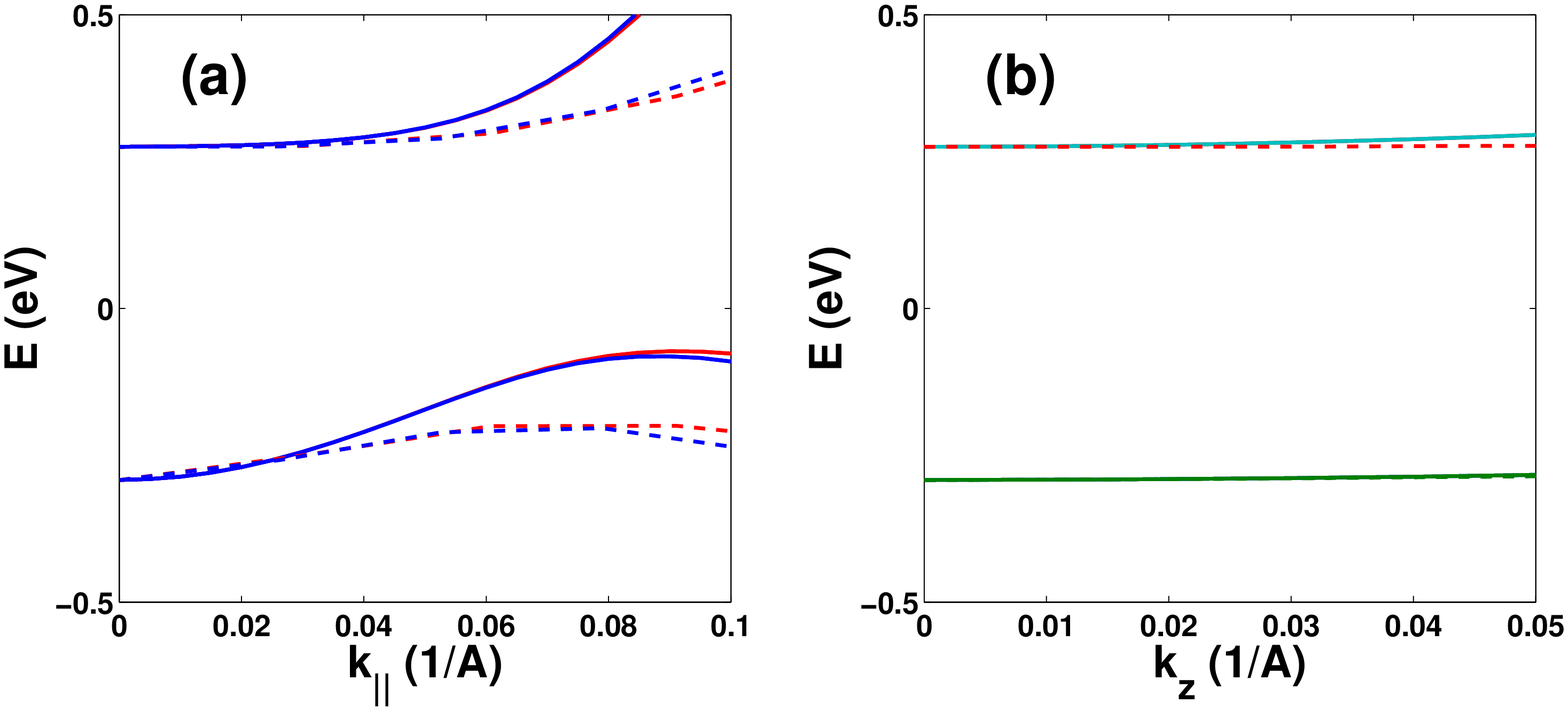}
      \includegraphics[width=3.5in]{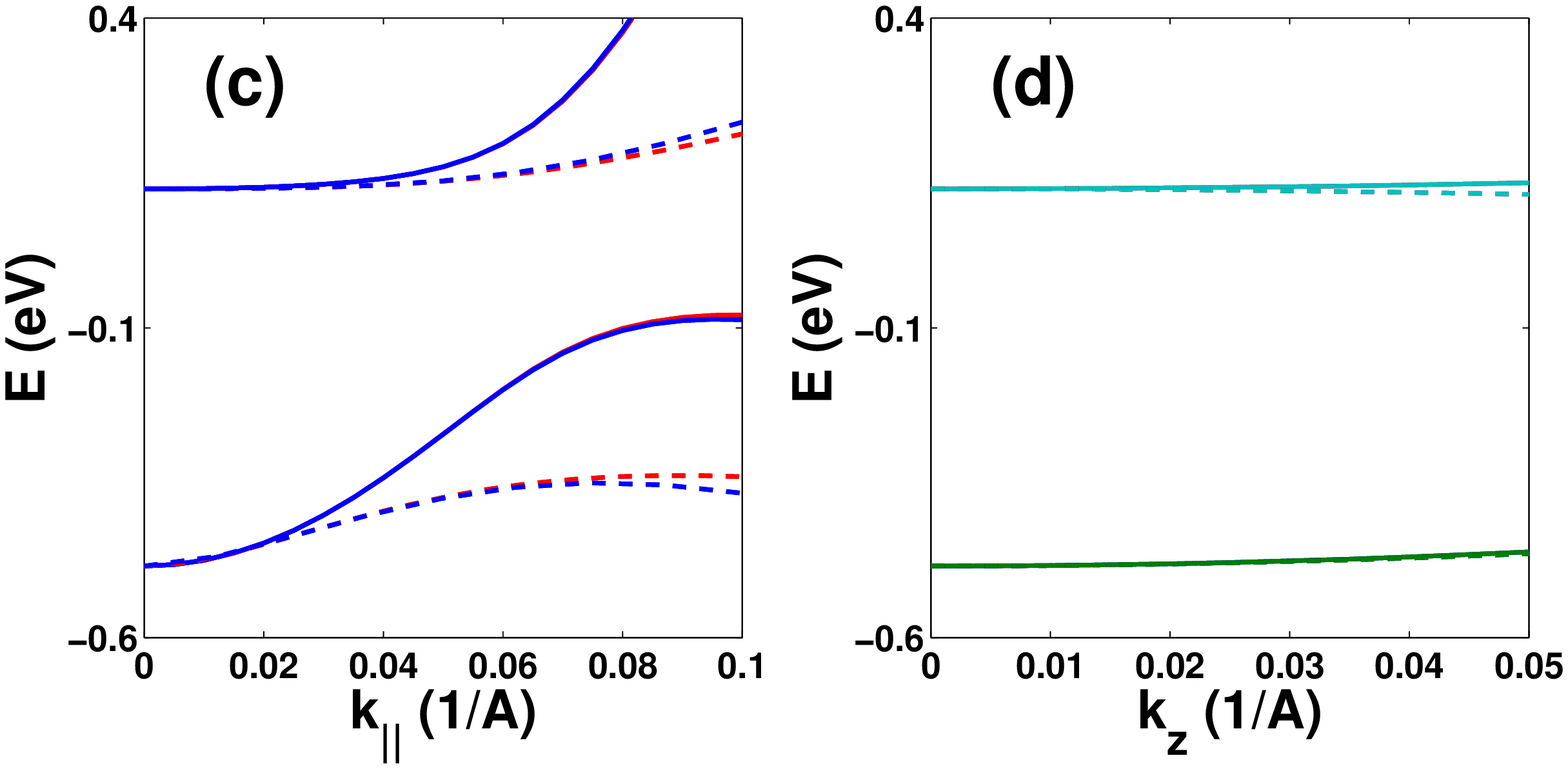}
      \includegraphics[width=3.5in]{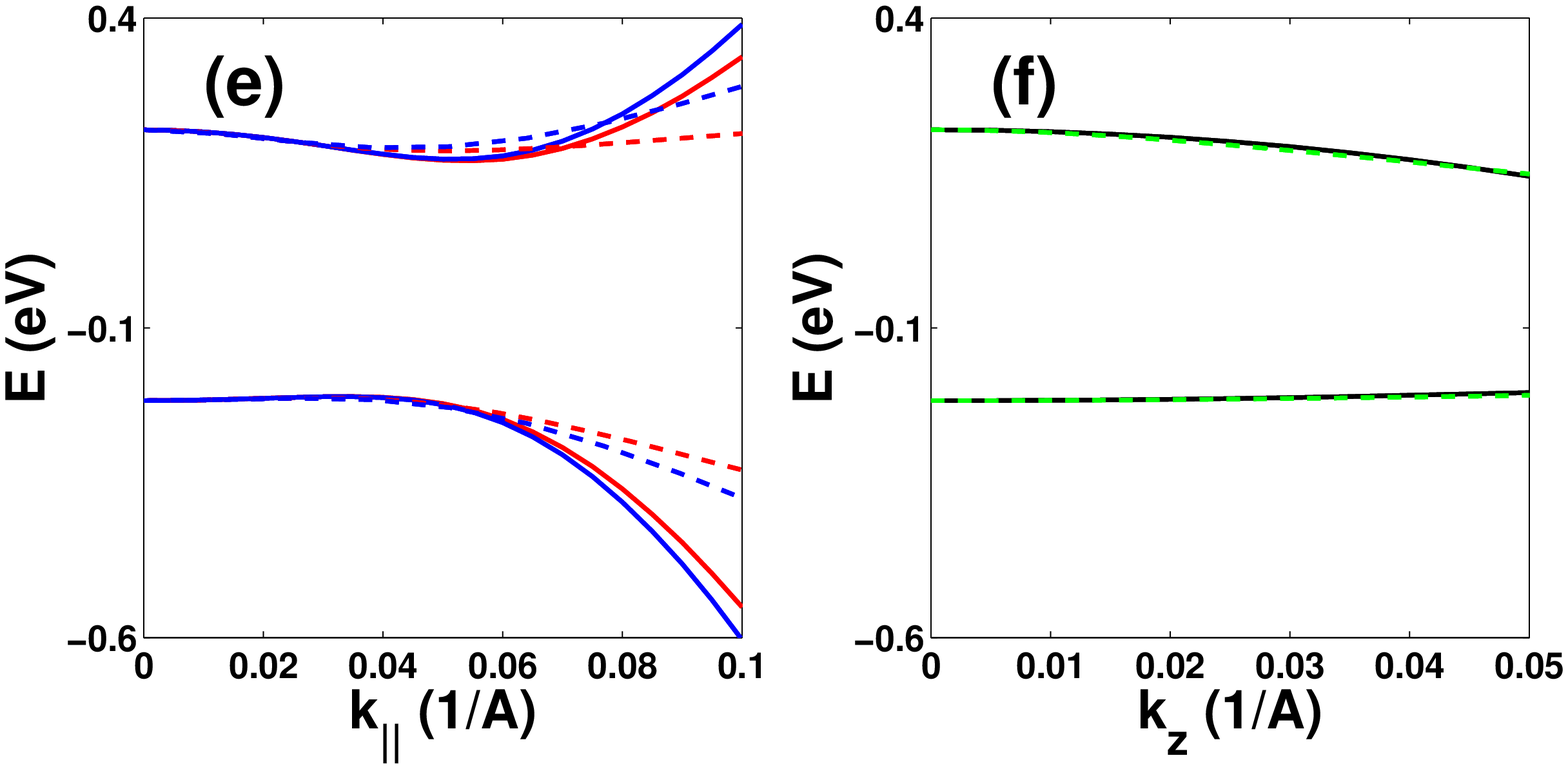}
    \end{center}
    \caption{ The energy dispersion obtained from our
    model Hamiltonian with four bands (solid line) is compared with that from
    {\it ab initio} calculation (dashed line). Here (a) and (b)
    is for $Bi_2Se_3$, (c) and (d) is for $Bi_2Te_3$, while
    (e) and (f) is for $Sb_2Te_3$. In (a), (c) and (e),
    the red line represents the dispersion along $k_x$ direction
    while the blue line is for $k_y$ direction. }
    \label{fig:dispersion}
\end{figure}

The key result of this section are the parameters of our model
Hamiltonian, given in table \ref{tab:4b_parameter}. The fitted
energy dispersions for $Bi_2Se_3$, $Bi_2Te_3$ and $Sb_2Te_3$ are
plotted in Fig \ref{fig:dispersion}. It is shown that our model
Hamiltonian is valid in the regime $k_{x,z}<0.04$\AA$^{-1}$.
However the maximum of the valence band for $Bi_2Se_3$ and
$Bi_2Te_3$ stays away from $\Gamma$ point, at about $k_x\approx0.07$\AA$^{-1}$,
therefore we need to keep in mind that there may be
some discrepancies when we try to use our model Hamiltonian to
describe $Bi_2Se_3$ quantitatively. Since our model Hamiltonian with
four bands already captures the salient features of the band
dispersion, especially the inverted band structure, in the following
two sections, we still stay in the framework of our model
Hamiltonian to discuss the topological surface states and the Landau
levels in the magnetic field. Then in the last section, we will
extend our model Hamiltonian to include eight bands, in order to
describe these materials more quantitatively.

\section{Surface states}
\label{sec:surface}
An important physical consequence of the non-trivial topology is the
existence of topological surface states. In this section, we would
like to study the surface state and its effective Hamiltonian based
on our model Hamiltonian derived above with open boundary condition.

Consider our model Hamiltonian (\ref{eq:Ham}) defined on the half
space given by $z>0$. We can divide our model Hamiltonian into two
parts
\begin{eqnarray}
    &&\hat{H}=\tilde{H}_0+\tilde{H}_1\\
    &&\tilde{H}_0=\tilde\epsilon(k_z)+\tilde{M}(k_z)\Gamma_5
    +B_0\Gamma_4k_z\\
    &&\tilde{H}_1=C_2k^2_{\parallel}+M_2k^2_{\parallel}\Gamma_5
    \nonumber\\
    &&+A_0(\Gamma_1k_y-\Gamma_2k_x)+H_3.
    \label{eq:sur_Ham}
\end{eqnarray}
where $\tilde\epsilon(k_z)=C_0+C_1k_z^2$ and $\tilde{M}(k_z)
=M_0+M_1k_z^2$. All $k_z$ dependent terms are included in
$\tilde{H}_0$. We replace $k_z$ by $-i\partial_z$, and obtain the
eigen-value equation
\begin{eqnarray}
\tilde{H}_0 (k_z \rightarrow -i\partial_z) \Psi(z)=E\Psi(z)
\end{eqnarray}
Since $\Gamma_4 =1\otimes\tau_2$ and $\Gamma_5=1\otimes\tau_3$ are
both block-diagonal, the Hamiltonian $\tilde{H}_0$ is also
block-diagonal and the eigen-states have the form
\begin{eqnarray}
        &&\Psi_\uparrow(z)=\left(
    \begin{array}{c}
        \psi_0\\
        {\bf 0}
    \end{array}
    \right)\qquad
    \Psi_\downarrow(z)=\left(
    \begin{array}{c}
    {\bf 0}\\
    \psi_0
    \end{array}
    \right),
    \label{eq:sur_wf}
\end{eqnarray}
where ${\bf 0}$ is a two-component zero vector.
$\Psi_\uparrow(z)$ is related to $\Psi_\downarrow(z)$ by time-reversal operation.
To obtain the surface states, the wave function $\psi_0(z)$ should
be localized at the surface and satisfies the eigen-equation
\begin{eqnarray}
    \left( \tilde\epsilon(-i\partial_z)+\tilde{M}(-i\partial_z)\tau_3
    -iB_0\tau_2\partial_z \right)\psi_0(z)=E\psi_0(z),
    \label{eq:sur_eqn}
\end{eqnarray}
which has been solved analytically for the open boundary
condition with different methods.\cite{koenig2008,zhou2008,shan2010,lu2009,linder2009}
Here in order to show the existence of the surface states and
to find the region where the surface states exist,
we would like to breifly review the derivation for
the explicit form of the surface states by neglecting
the $\tilde\epsilon$ for simplicity\cite{koenig2008}.

After neglecting the $\tilde\epsilon$ term, the eigen equation
(\ref{eq:sur_eqn}) has the particle hole symmetry, therefore
we would like to expect a sepecial surface states with $E=0$
can exist. With the wave function ansatz $\psi_0=\phi e^{\lambda z}$,
the above equation can be simplified as
\begin{eqnarray}
    \left( M_0-M_1\lambda^2 \right)
    \tau_1\phi=B_0\lambda\phi.
    \label{eq:sur_eqn_cont1}
\end{eqnarray}
It is obvious that the two-component wave function $\phi$ should be the eigen-state
of Pauli matrix $\tau_1$. Let's define $\tau_1\phi_\pm=\pm\phi_\pm$, then the equation
(\ref{eq:sur_eqn_cont1}) is simplified to a quadratic equation for $\lambda$.
Another important observation is that if $\lambda$ is a solution for $\phi_+$,
then $-\lambda$ is the solution for $\phi_-$. Consequently, the generic
wave function is given by
\begin{eqnarray}
    \psi_0(z)=(ae^{\lambda_1 z}+be^{\lambda_2 z})\phi_++(ce^{-\lambda_1 z}+de^{-\lambda_2 z})\phi_-,
    \label{eq:sur_wf1}
\end{eqnarray}
where $\lambda_{1,2}$ satisfy
\begin{eqnarray}
    \lambda_{1,2}=\frac{1}{2M_1}\left( -B_0\pm\sqrt{4M_0M_1+B_0^2 }\right).
    \label{eq:sur_lambda1}
\end{eqnarray}
Similar to the Ref. [\onlinecite{koenig2008}], the open boundary condition
$\psi(0)=0$, together with the normalizability of the wave function in the region
$z>0$, leads to the existence condition of the surface states,
$\Re\lambda_{1,2}<0$ ($c=d=0$) or $\Re\lambda_{1,2}>0$ ($a=b=0$),
which can only be satisfied with the band inversion condition $M_0M_1<0$. Furthermore, it is easy to show that when $B_0/M_1>0$, $\Re\lambda_{1,2}<0$,
while $B_0/M_1<0$, $\Re\lambda_{1,2}>0$, thus the wave function for the surface states at
$\Gamma$ point is given by
\begin{eqnarray}
    \psi_0(m)=\left\{
    \begin{array}{cc}
        a\left( e^{\lambda_1 z}-e^{\lambda_2 z} \right)\phi_+&B_0/M_1>0\\
        c\left( e^{-\lambda_1z}-e^{-\lambda_2z} \right)\phi_-&B_0/M_1<0
    \end{array}
    \right. .
    \label{eq:sur_wfpsi}
\end{eqnarray}
We emphasize here the sign change of $B_0/M_1$ will change the spin basis of the surface states, which
will be the key point to determine the helicity of the Dirac Hamiltonian for the topological surface states.
Another important quantity of the surface states is the decaying length, which can be defined
as $l_c={\rm max}\left\{ \frac{1}{|\Re(\lambda_{1,2})|} \right\}$, given by
\begin{eqnarray}
    l_c=\left\{
    \begin{array}{cc}
    \Re\left( \frac{B_0+\sqrt{4M_0M_1+B_0^2}}{2M_0} \right)&B_0>0,M_1<0\\
    \Re\left( \frac{B_0-\sqrt{4M_0M_1+B_0^2}}{2M_0} \right)&B_0<0,M_1>0\\
    \Re\left( -\frac{B_0+\sqrt{4M_0M_1+B_0^2}}{2M_0} \right)&B_0>0,M_1>0\\
    \Re\left( -\frac{B_0-\sqrt{4M_0M_1+B_0^2}}{2M_0} \right)&B_0<0,M_1<0\\
    \end{array}
    \right.,
    \label{eq:sur_lc}
\end{eqnarray}
where $\Re$ takes the real part.

In the above, we take a simple tight-binding model to show the existence condition
and the form of the wave function for the surface states, which
can help us to understand the underlining physics qualitively, but not
quantitively. In the realistic materials, the detail form of $\psi_0$ will
depend on the material detail, such as the boundary condition or the detail
parameters, however the form of the wave function (\ref{eq:sur_wf})
remains valid. Therefore in the following, we just simply treat
$\psi_0$ by some parameters. On the sub-space $\Psi=[
\Psi_\uparrow,\Psi_\downarrow]$, we find that
\begin{eqnarray}
    &&\langle \Psi|\Gamma_1|\Psi\rangle=\alpha_1\sigma_x,\qquad
        \langle \Psi|\Gamma_2|\Psi\rangle=\alpha_1\sigma_y,\nonumber\\
        &&\langle \Psi|\Gamma_3|\Psi\rangle=\alpha_1\sigma_z,\qquad
        \langle \Psi|\Gamma_4|\Psi\rangle=0,\nonumber\\
        &&\langle \Psi|\Gamma_5|\Psi\rangle=\alpha_3,
\end{eqnarray}
with $\alpha_1\equiv\langle \psi_0|\tau_1|\psi_0\rangle$
and $\alpha_3\equiv\langle \psi_0|\tau_3|\psi_0\rangle$.
With these expressions, the effective Hamiltonian of the surface
states $\Psi$ is given by\cite{zhang2009}
\begin{eqnarray}
    &&H_{sur}=\tilde{C}_0+\tilde{C}_2k_{\parallel}^2+\tilde{A}
    (\sigma_xk_y-\sigma_yk_x)+\tilde{R}(k_+^3+k_-^3)\sigma_z
    \nonumber\\&&=(C_0+\alpha_3M_0)+(C_2+\alpha_3M_2)
    k_{\parallel}^2+A_0\alpha_1(\sigma_xk_y-\sigma_yk_x)\nonumber\\
    &&+\frac{R_1\alpha_1}{2}(k_+^3+k_-^3)\sigma_z
    \label{eq:sur_Heff}
\end{eqnarray}
with $k_\pm=k_x\pm ik_y=k_\parallel e^{\pm i\theta}$. The $k^3$
terms have also been found in Ref. [\onlinecite{fu2009}].
In the following numerical calculation, the coefficient $\alpha_1$
and $\alpha_3$ are treated as two fitting parameters to the
experiment, given by $\alpha_1=\frac{\tilde{A}_{exp}}{A_0}=0.99$ and
$\alpha_3=\frac{\tilde{C}_{exp}-C_0}{M_0}=-0.15$, where $\tilde{A}_{exp}
=3.29eV\cdot\AA$ comes from the Fermi velocity of the surfaces states
and $\tilde{C}_{exp}=0.035eV$ comes from the position of the surface
Dirac points\cite{xia2009}. Further we need to
check the spin operators in this system. Again we use the wave
function from {\it ab initio} calculation and project the spin
operator into the subspace spanned by the four basis states. After
we obtain the spin operators for our model Hamiltonian, we can use
the eigen wave function (\ref{eq:sur_wf}) to project the spin
operator into the surface states subspace. Finally we find that
$\langle \Psi|S_x|\Psi\rangle=S_{x0}\sigma_x$, $\langle
\Psi|S_y|\Psi\rangle=S_{y0}\sigma_y$ and $\langle
\Psi|S_z|\Psi\rangle=S_{z0}\sigma_z$ with $S_{x(y,z)0}$ to be some
constants.
This indicates that $\sigma$ matrix in our model Hamiltonian
(\ref{eq:sur_Heff}) is proportional to the real spin.

\begin{figure}[htpb]
   \begin{center}
      \includegraphics[width=3.5in]{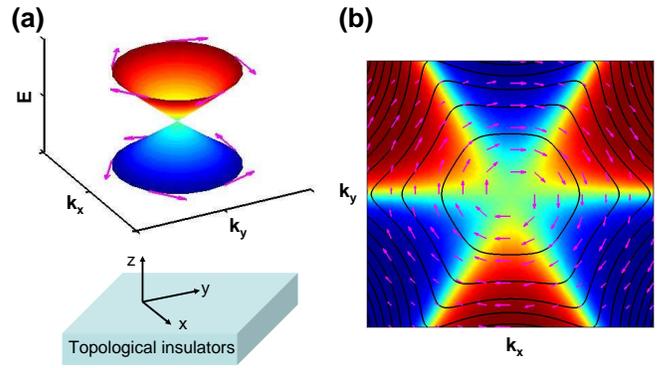}
    \end{center}
    \caption{ (a) Spin texture of the surface states near $\Gamma$ point.
    For conduction band, the helicity is left handed while for valence band, it is right handed.  (b) Spin texture of the conduction band of the surface states
    in the momentum space.
    The arrow represents the x-y planar spin polarization while the color indicates
    the z component of the spin polarization. Here red is for spin
    up while blue is for spin down. The black line
    gives the constant energy contours. }
    \label{fig:surface}
\end{figure}

The derivation of the surface Hamiltonian (\ref{eq:sur_Heff}) is the
central result of this section.
In the limit $k\rightarrow 0$, the linear term in Hamiltonian (\ref{eq:sur_Heff})
will be dominant, then the surface states
show the linear dispersion with helical spin texture, which has
the opposite direction for the conduction and valence band, as shown in Fig.\ref{fig:surface} (a).
Such type of spin texture is similar to one of the fermi surfaces
in the usual 2D electron gas with Rashba SOC\cite{bychkov1984,winklerbook2003},
which can be simply understood from the
fact that the inversion symmetry is broken near the surface.
The helical spin texture has also been calculated from {\it ab initio}
method\cite{zhang2010} and already observed in the pioneering spin-resolved
APRES measurement\cite{hsieh2009}.
From (\ref{eq:sur_Heff}), the helicity of the spin texture is determined
by the sign of the coefficient $\tilde{A}=A_0\alpha_1$,
where $\alpha_1\equiv\langle \psi_0|\tau_1|\psi_0\rangle$ is related
to the spin basis of the surface states (\ref{eq:sur_wfpsi}).
Therefore the helicity is determined by the relative sign of $A_0$
and $B_0/M_1$. Furthermore, due to the inversion condition $M_1M_0<0$,
the sign of $M_1$ is already determined by the gap $M_0$.
Consequently, within our model Hamiltonian the helicity of the spin texture
is given by the relative sign of the coefficients of two types of linear terms,
$A_0$ and $B_0$.

To further explore the origin of the helicity of the spin texture
in the atomic levels, we relate the coefficients $A_0$ and $B_0$
to the atomic SOC by using the expression (\ref{eq:struct_SOCwf3})$\sim$(\ref{eq:struct_SOCwf6}),
as
\begin{eqnarray}
    &&A_0=\frac{\hbar}{2m_0}\langle P1^+_-,\frac{1}{2}|p_+|P2^-_+,-\frac{1}{2}\rangle\nonumber\\
    &&=\frac{\hbar}{2m_0}\left[ \left( u^{P1^+}_-v^{P2^-}_+ \right)^*\langle P1^+,p_z|p_+|P2^-,p_-\rangle\right.\nonumber\\
    &&\left.+\left( v^{P1^+}_-u^{P2^-}_+ \right)^*\langle P1^+,p_+|p_+|P2^-,p_z\rangle\right]\\
    &&B_0=\frac{\hbar}{m_0}\langle P1^+_z,p_z|p_z|P2^-,p_z\rangle\nonumber\\
    &&=\frac{\hbar}{m_0}\left[ (u^{P1^+}_-)^*u^{P2^-}_+\langle P1^+,p_z|p_z|p2^-,p_z\rangle\right.\nonumber\\
    &&\left. +(v^{P1^+}_-)^*v^{P2^-}_+\langle P1^+,p_+|p_z|P2^-,p_+\rangle\right].
    \label{eq:sur_A0B0}
\end{eqnarray}
Here $|\Lambda,\alpha\rangle$ ($\Lambda=P1^+,P2^-$ and $\alpha=p_x,p_y,p_z$)
are the atomic orbitals without any SOC
and all the dependence of SOC are included in the coefficient $u^\Lambda$ and $v^\Lambda$.
From (\ref{eq:struct_uv}), we find that for $u^{\Lambda_1}v^{\Lambda_2}$ it is
proportional to $\lambda_{\Lambda_1}$ (or $\lambda_{\Lambda_2}$),
which indicates that $A_0$ depends on the sign of the atomic SOC, while
it is only possible for $(u^{\Lambda_1})^*u^{\Lambda_2}$ and $(v^{\Lambda_1})^*v^{\Lambda_2}$
to be independent of $\lambda_{\Lambda_{1,2}}$ or be proportional to
$\lambda_{\Lambda_1}\lambda_{\Lambda_2}$, thus the sign of $B_0$
will not depend on the atomic SOC. Finally, we conclude that the helicity of the spin texture is
originally related to the atomic SOC.

In the above, we have shown how the the linear term determines the spin texture of
the surface states, which can also be affected by the quadratic term and cubic term
in the effective Hamiltonian (\ref{eq:sur_Heff}). We solve the eigenvalue problems of
the whole effective Hamiltonian and obtain the eigen-energy and eigen states as
\begin{eqnarray}
    &&E_\pm=\tilde{C}_0+\tilde{C}_2k^2_\parallel
    \pm\sqrt{\tilde{A}^2k^2_{\parallel}+4\tilde{R}^2k^6\cos^23\theta}\\
    &&\psi_\pm=\frac{1}{\sqrt{N}}\left(
    \begin{array}{c}
        \tilde{A}(k_y+ik_x)\\
        d_\pm-\tilde{R}(k_+^3+k_-^3)
    \end{array}
    \right)
    \label{eq:sur_eigsur}
\end{eqnarray}
with $d_\pm=\pm\sqrt{\tilde{A}^2k^2_{\parallel}+4\tilde{R}^2k^6\cos^23\theta}$ and
$N=\tilde{A}^2k^2+(\sqrt{\tilde{A}^2k^2+4\tilde{R}^2k^6\cos^23\theta}-2\tilde{R}k^3\cos3\theta)^2$.
Consequently the spin polarization in k space is given by
\begin{eqnarray}
    &&\langle \psi_+|\sigma_x|\psi_+\rangle=\frac{2\tilde{A}k_y}{N}(d_+-2\tilde{R}k^3\cos3\theta)\\
        &&\langle \psi_+|\sigma_y|\psi_+\rangle=-\frac{2\tilde{A}k_x}{N}(d_+-2\tilde{R}k^3\cos3\theta)\\
    &&\langle \psi_+|\sigma_z|\psi_+\rangle=\frac{4\tilde{R}k^3\cos3\theta}{N}(
    d_+-2\tilde{R}k^3\cos3\theta)
\end{eqnarray}
which is plotted in Fig \ref{fig:surface} (b).
In the limit $k\rightarrow0$, the spin polarization almost lies
in the $xy$ plane, which is due to the linear term and has been
discussed in the above. When $k$ is increased, the k-cubic term
comes into play, which will not only induce the hexagonal warping of the
constant energy contours\cite{fu2009} but also yield z direction
spin polarization, similar to the situation in $Bi_xSb_{1-x}$
studied by {\it ab initio} calculation\cite{zhang2009d}.

\section{Magnetic field and Landau level}
\label{sec:LL}

In this section, we study the Landau level problem for both the bulk
states and the surface states, which is important for predicting or
understanding many properties of the system in a magnetic field,
such as the SdH oscillation, surface quantum Hall effect and
magneto-optics. In this regards, our model Hamiltonian has the
unique advantage, since the magnetic field effect cannot be
incorporated in {\it ab initio} calculations. For the realistic
finite sample, the bulk Landau levels will always coexist with the
surface Landau levels, thus both the two types of Landau levels need
to be taken into account. For simplicity, we solve the bulk Landau
level for an infinite sample and the surface Landau level for the
semi-infinite sample. The mixing effect between bulk and surface
Landau levels is neglected here.

\begin{figure}[htpb]
    \vspace{0.3in}
    \begin{center}
      \includegraphics[clip,width=2.3in]{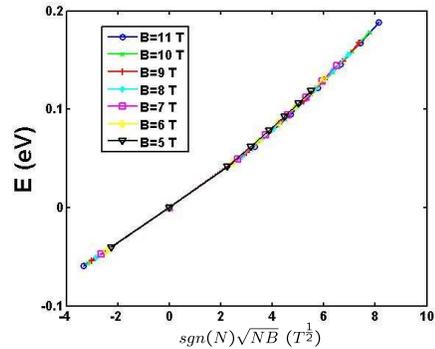}
    \end{center}
    \caption{ The energy of the Landau level verse $sgn(N)\sqrt{NB}$ where
    $N$ is the Landau level index and $B$ is the magnetic field.  }
    \label{fig:LanInd}
\end{figure}

\begin{figure}[htpb]
   \begin{center}
      \includegraphics[width=3.3in]{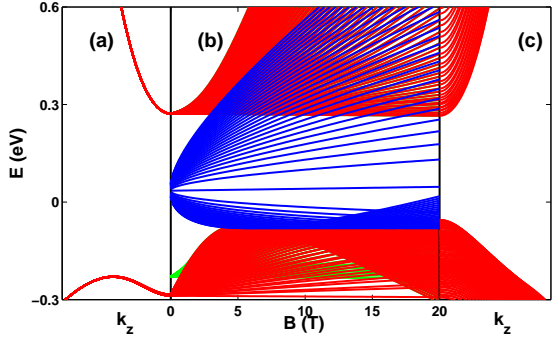}
    \end{center}
    \caption{ The Landau levels in the magnetic field for both
    the bulk states (red lines) and the surface states (blue lines)
    are shown in (b). Also the dispersion along z direction at (a)
    $B=0T$ and (c) $B=20T$ is plotted. }
    \label{fig:landau}
\end{figure}

For the bulk state, there are two types of contribution from the
magnetic field, the orbital effect and the Zeeman effect. The
orbital effect can be included by Peierls
substitution\cite{luttinger1955} ${\bf k}\rightarrow
{\bf \pi}={\bf k}+\frac{e}{\hbar}{\bf A}$ with
${\bf A}=(0,B_zx,0)$ for magnetic field along z-direction. We
introduce the annihilation and creator operators
$a=\frac{l_c}{\sqrt{2}}\pi_-$ and $a^\dag=\frac{l_c}{\sqrt{2}}\pi_+$
with $l_c=\sqrt{\frac{\hbar}{eB_z}}$ for the harmonic oscillator
function $\varphi_n$. $a$ and $a^\dag$ satify $a\varphi_N=
\sqrt{N}\varphi_{N-1}$, $a^{\dag}\varphi_N=\sqrt{N+1}\varphi_{N+1}$
and $[a,a^\dag]=1$. With the operators $a$ and $a^\dag$, the
Hamiltonian (\ref{eq:Ham0}) is written as
\begin{eqnarray}
    \hat{H}_{0B}=\tilde\epsilon+\left(
    \begin{array}{cccc}
        \tilde\mathcal{M}&B_0k_z&0&A_0\frac{\sqrt{2}}{l_c}a\\
        B_0k_z&-\tilde\mathcal{M}&A_0\frac{\sqrt{2}}{l_c}a&0\\
        0&A_0\frac{\sqrt{2}}{l_c}a^\dag&\tilde\mathcal{M}&-B_0k_z\\
        A_0\frac{\sqrt{2}}{l_c}a^\dag&0&-B_0k_z&-\tilde\mathcal{M}
    \end{array}
    \right)\label{eq:LL_Ham1}
\end{eqnarray}
where $\tilde\epsilon(k_z,a^\dag a)=C_0+C_1k_z^2+\frac{2C_2}{l_c^2}
\left( a^\dag a+\frac{1}{2} \right)$ and $\tilde{M}(k_z,a^\dag
a)=M_0+M_1k_z^2+\frac{2M_2}{l_c^2} \left( a^\dag a+\frac{1}{2}
\right)$. The $k^3$ term, which breaks the in-plane rotation
symmetry, is neglected here, therefore the wave function should have
the form of $\Psi_N=[f^N_1\varphi_{N-1},f^N_2\varphi_{N-1},
f^N_3\varphi_N,f^N_4\varphi_N]^T$. With such wave function ansatz,
the Hamiltonian is transformed to
\begin{eqnarray}
    \hat{H}_{0B}(k_z,N)=\left(
    \begin{array}{cccc}
        \tilde\mathcal{M}^+_{N-1}&B_0k_z&0&A_0\frac{\sqrt{2N}}{l_c}\\
        B_0k_z&\tilde\mathcal{M}^-_{N-1}&A_0\frac{\sqrt{2N}}{l_c}&0\\
        0&A_0\frac{\sqrt{2N}}{l_c}&\tilde\mathcal{M}^+_N&-B_0k_z\\
        A_0\frac{\sqrt{2N}}{l_c}&0&-B_0k_z&-\tilde\mathcal{M}^-_N
    \end{array}
    \right)\label{eq:LL_Ham2}
\end{eqnarray}
with $\tilde\mathcal{M}^+_N=\tilde\epsilon(k_z,N)+\tilde\epsilon(k_z,N)$
and $\tilde\mathcal{M}^-_N=\tilde\epsilon(k_z,N)-\tilde\epsilon(k_z,N)$.
%

To consider the Zeeman splitting, we need to further calculate the
effective g-factor\cite{winklerbook2003,roth1959}. The effective
Zeeman type coupling can also written down for our model Hamiltonian
by symmetry principles. By a quick inspection of the table
(\ref{tab:Character3}), we find that the following terms
\begin{eqnarray}
    &&\hat{H}_Z=\tilde{g}_{z1}\Gamma_{12}B_z+\tilde{g}_{z2}\Gamma_{34}B_z\nonumber\\
    &&+\tilde{g}_{xy1}\left(
    \begin{array}{cc}
        \Gamma_{23}&\Gamma_{31}
    \end{array}
    \right)\left(
    \begin{array}{cc}
        \cos\phi_1&\sin\phi_1\\
        -\sin\phi_1&\cos\phi_1
    \end{array}
    \right)\left(
    \begin{array}{c}
        B_x\\B_y
    \end{array}
    \right)\nonumber\\
    &&+\tilde{g}_{xy2}\left(
    \begin{array}{cc}
        \Gamma_{14}&\Gamma_{24}
    \end{array}
    \right)\left(
    \begin{array}{cc}
        \cos\phi_2&\sin\phi_2\\
        -\sin\phi_2&\cos\phi_2
    \end{array}
    \right)\left(
    \begin{array}{c}
        B_x\\B_y
    \end{array}
    \right)
    \label{eq:kp_Hmag1}
\end{eqnarray}
are possible couplings to the magnetic field. Again $\phi_1$ and
$\phi_2$ are phase factors which need to be determined from other
methods, and here we take $\phi_1=\phi_2=0$ to coincide with the
results from the ${\bf k}\cdot{\bf p}$ method, and explicitly
(\ref{eq:kp_Hmag1}) can be written as
\begin{eqnarray}
    \hat{H}_Z=\frac{\mu_B}{2}\left(
    \begin{array}{cccc}
        g_{1z}B_z&0&g_{1p}B_-&0\\
        0&g_{2z}B_z&0&g_{2p}B_-\\
        g_{1p}B_+&0&-g_{1z}B_z&0\\
        0&g_{2p}B_+&0&-g_{2z}B_z
    \end{array}
    \right)
    \label{eq:kp_Hmag2}
\end{eqnarray}
with $\mu_B=\frac{e\hbar}{2m_0}$,  and
$\tilde{g}_{xy1}+\tilde{g}_{xy2}=\frac{\mu_B}{2}g_{1p}$,
$\tilde{g}_{xy1}-\tilde{g}_{xy2}=\frac{\mu_B}{2}g_{2p}$,
$\tilde{g}_{z1}+\tilde{g}_{z2}=\frac{\mu_B}{2}g_{1z}$,
$\tilde{g}_{z1}-\tilde{g}_{z2}=\frac{\mu_B}{2}g_{2z}$. This model
Hamiltonian can also be derived from ${\bf k}\cdot {\bf p}$ theory
and the parameters $g_{1z}$, $g_{2z}$, $g_{1p}$ and $g_{2p}$ can be
related to the matrix elements of the momentum operator $\bf p$ in
the ${\bf k}\cdot{\bf p}$ theory, with full details given in
Appendix \ref{app:C_kp}. Now our total Hamiltonian for the bulk
states under the z-direction magnetic field is given by
$\hat{H}_B=\hat{H}_{0B}+\hat{H}_Z$,
which can be solved numerically to obtain
the Landau level $E^{bulk}_{N,\eta}(B,k_z)$ with Landau level index $N$
and band index $\eta$ under the z-direction magnetic field.

Similar procedure can be applied to the surface effective
Hamiltonian (\ref{eq:sur_Heff}). With the wave function ansatz
$\Psi_{sur,N}=[g^N_1\varphi_{N-1}, g^N_2\varphi_N]^T$, the surface
Hamiltonian is changed to
\begin{eqnarray}
    &&\hat{H}_{sur,B0}(N)=\tilde{C}_0+\frac{2NeB_z}{\hbar}\tilde{C}_2-
    \frac{\tilde{C}_2eB_z}{\hbar}\sigma_z\nonumber\\
    &&-\sqrt{\frac{2NeB_z}{\hbar}}\tilde{A}\sigma_y,
    \label{eq:ll_sur}
\end{eqnarray}
and the Zeeman type term is given by
\begin{eqnarray}
    \hat{H}_{sur,Z}=\frac{\mu_B}{2}g_{sz}\sigma_zB_z+\frac{\mu_B}{2}g_{sp}\left(
    \sigma_xB_x+\sigma_yB_y\right)
    \label{eq:kp_HsurB}
\end{eqnarray}
with $\frac{\mu_B}{2}g_{sz}=\tilde{g}_{z1}+\tilde{g}_{z2}\alpha_3$ and
$\frac{\mu_B}{2}g_{sp}=\tilde{g}_{xy1}+\tilde{g}_{xy2}\alpha_3$.
The total Hamiltonian for the surface states yields $\hat{H}_{sur,B}
=\hat{H}_{sur,B0}+\hat{H}_{sur,Z}$ and correspondingly the Landau
level in z-direction magnetic field $B_z$ is solved as
\begin{eqnarray}
    &E^{sur}_{s}(N)=\tilde{C}_0+\frac{2NeB_z}{\hbar}\tilde{C}_2+\nonumber\\
    &s\sqrt{\left( -\frac{\tilde{C}_2eB_z}{\hbar}+\frac{\mu_B}{2}g_{sz}B_z \right)^2
    +\frac{2NeB_z}{\hbar}\tilde{A}^2}
    \label{eq:LL_eigsur}
\end{eqnarray}
with $s=\pm$ for $N=1,2,\cdots$ and
\begin{eqnarray}
    &E^{sur}(0)=\tilde{C}_0+\frac{eB_z}{\hbar}\tilde{C}_2-\frac{\mu_B}{2}g_{sz}B_z
    \label{eq:LL_eigsur0}
\end{eqnarray}
for zero mode $N=0$.
Here we note that due to the existence of the quadratic term
$\tilde{C}_2k^2_{\parallel}$, the square root dependence of the
energy level verse magnetic field is only an approximation
applicable for low Landau levels and low magnetic field. For high
magnetic field, it will be a combination of the linear contribution
and square root contribution. As shown in Fig \ref{fig:LanInd}, the
energy of the Landau levels are plotted as the function of
$sgn(N)\sqrt{NB_z}$ and the non-linear behavior will appear for high
$\sqrt{NB_z}$.

\begin{figure}[htpb]
   \begin{center}
      \includegraphics[width=3in]{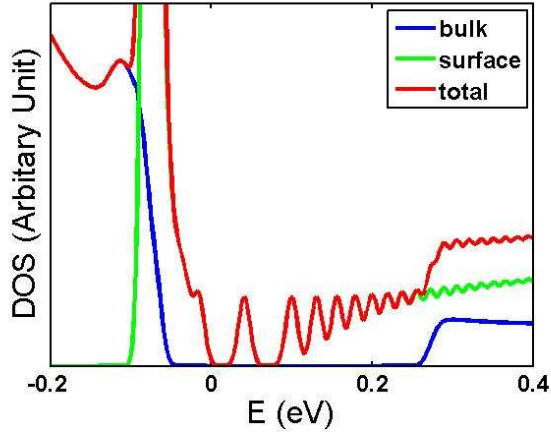}
    \end{center}
    \caption{ The density of states (DOS) as a function of energy is plotted
    with $B=10T$ for the bulk (blue line), surface (green line)
    and total (red line) Landau levels.}
    \label{fig:dos}
\end{figure}

In Fig \ref{fig:landau} (b), the Landau levels for both bulk states
and surface states are plotted as a function of magnetic field.
Here we emphasize that for the bulk states,
the Landau levels are plotted for the $k_z=0$ point
(red line in Fig \ref{fig:landau} (b)). The dispersion along z
direction is also shown in Fig \ref{fig:landau} (a) for $B=0T$ and
(c) for $B=20T$. From Fig \ref{fig:landau} (a), we find that
the maximum of the valence band is not located at $k_z=0$ point
for small magnetic field, hence we plot the maximum of valence
band as the green lines in Fig \ref{fig:landau} (b).
When the magnetic field is increased,
the bulk gap is also decreased by a significant amount
(about 160meV for 5T magnetic field), which is due to the
double hump structure for the valence band dispersion of $Bi_2Se_3$.
Such decrease may be observed in a magneto-optical measurement.

In order to compare with the scanning tunneling microscope (STM)
experiment\cite{cheng2010,hanaguri2010},
it is helpful to investigate the local density of states (LDOS) at
the surface. The LDOS for the surface states and the bulk states can
be obtained by\cite{winklerbook2003}
\begin{eqnarray}
    D_{sur}(E,B)=\sum_{N,s}\frac{G}{\sqrt{2\pi}\Gamma}
    e^{-\frac{(E-E^{sur}_{N,s}(B))^2}{2\Gamma^2}}
    \label{eq:LL_dossur}
\end{eqnarray}
and
\begin{eqnarray}
    D_{bulk}(E,B)=L_0\int\frac{dk_z}{2\pi}\sum_{N,s}\frac{G}{\sqrt{2\pi}\Gamma}
    e^{-\frac{(E-E^{bulk}_{N,\eta}(B,k_z))^2}{2\Gamma^2}}
    \label{eq:LL_dosbulk}
\end{eqnarray}
respectively, where $G=\frac{eB_z}{2\pi\hbar}=\frac{1}{2\pi l^2_c}$
is the degeneracy of each Landau level and $\Gamma$ is the broading.
In order to compare the bulk LDOS with the surface LDOS, we require
to introduce a length scale $L_0$ which represents the detection
depth of STM. Here we simply take $L_0$ to be the thickness of one
quintuple layer. Furthermore the surface states only exist near
${\bf k}=0$, thus we need to take a cut-off for the Landau level
index $N$. With the formula (\ref{eq:LL_dossur}) and
(\ref{eq:LL_dosbulk}), LDOS for both the bulk and surface Landau
levels are shown in Fig. \ref{fig:dos}. The bulk LDOS shows a gap of
about 0.3eV and within the bulk gap, only surface LDOS remains and
shows clearly the Landau levels as discreted peaks. The largest
Landau gap for surface states is between the 0th and 1th Landau
level, about 50meV, which is large enough for the observation of the
topological magneto-electric effect\cite{qi2008b,qi2009}.

\begin{figure}[htpb]
   \begin{center}
      \includegraphics[width=3.5in]{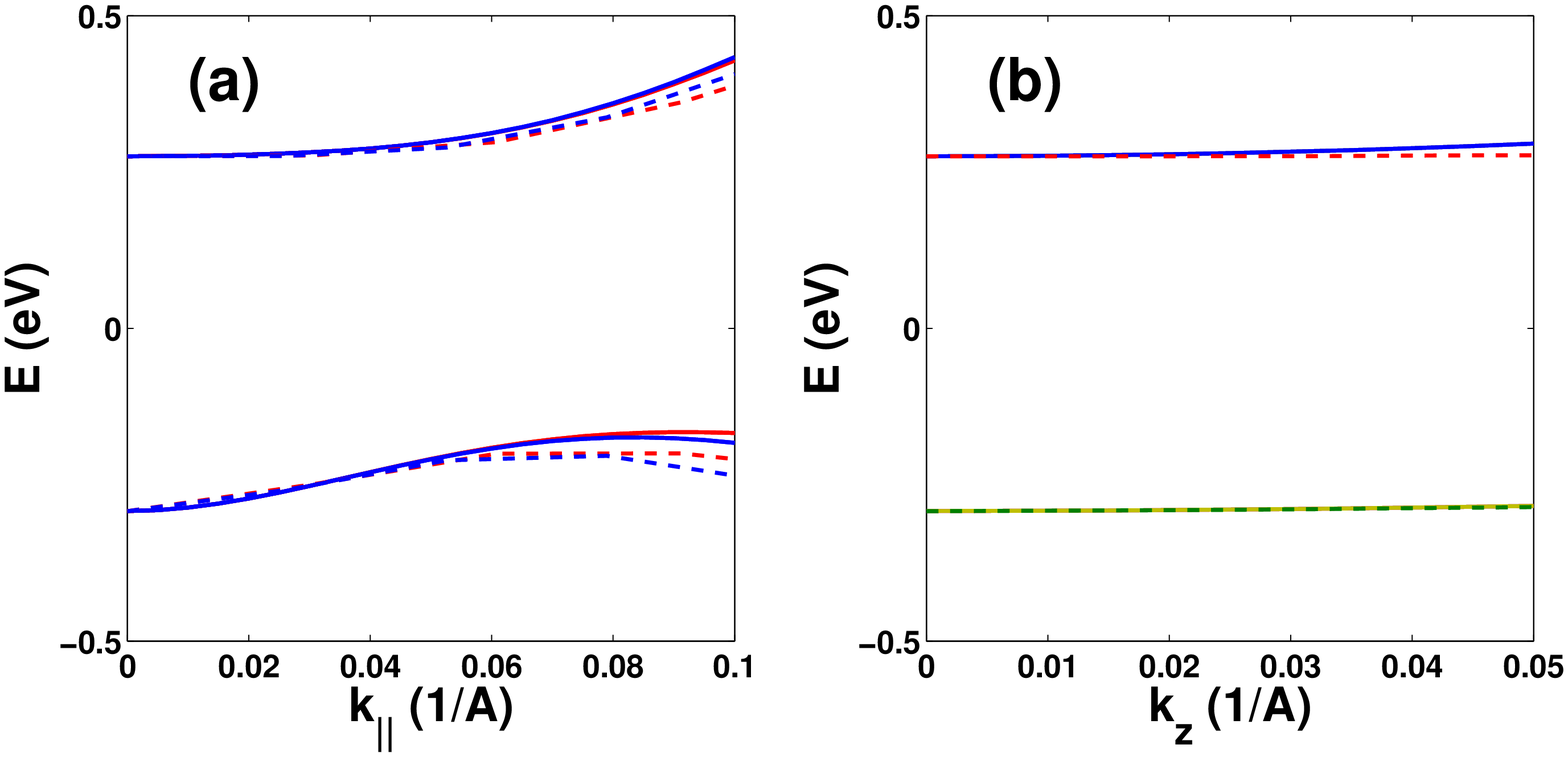}
      \includegraphics[width=3.5in]{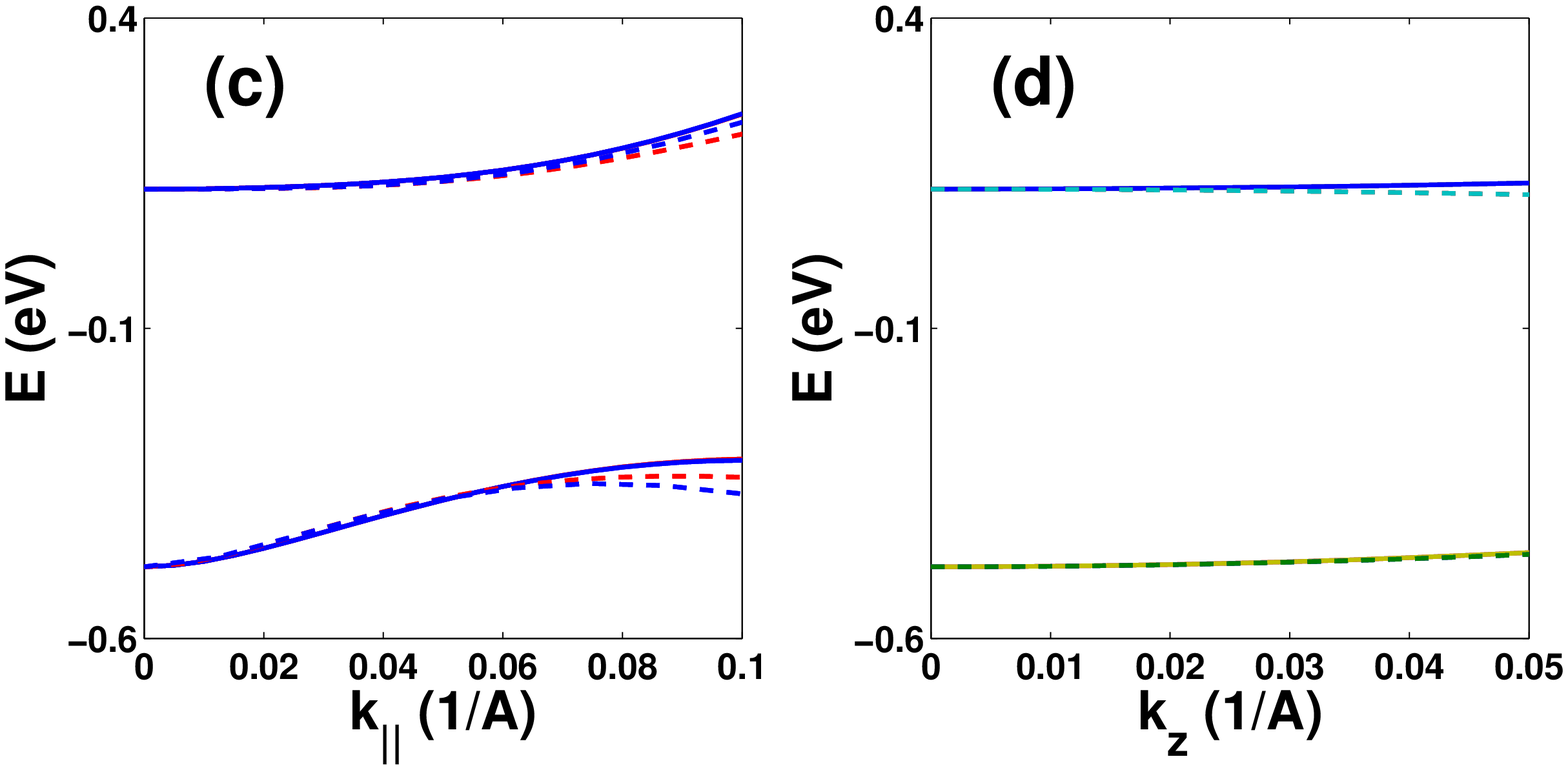}
      \includegraphics[width=3.5in]{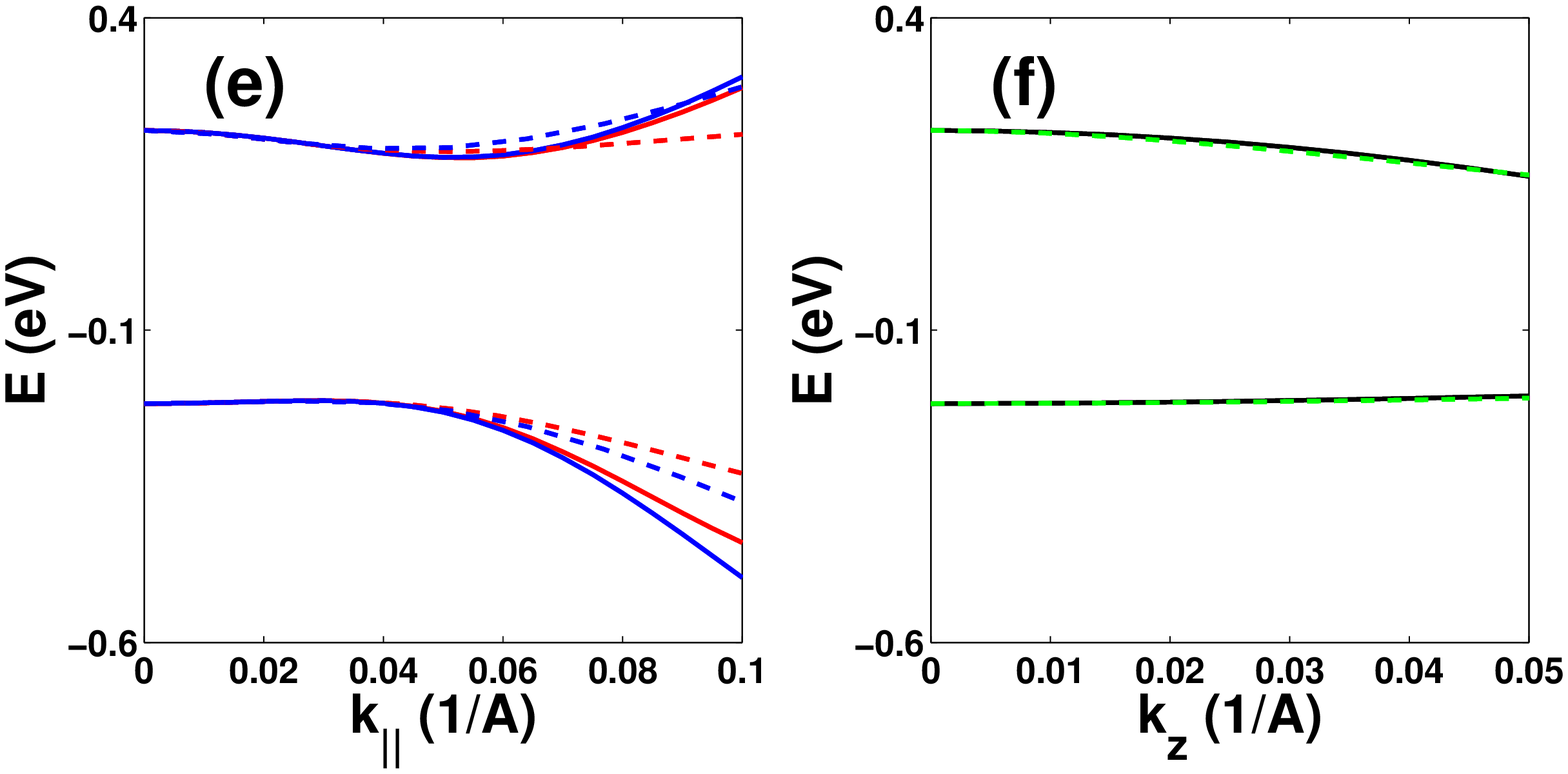}
    \end{center}
    \caption{ The energy dispersion obtained from the new model
Hamiltonian with eight bands (solid line) is compared with that from
    {\it ab initio} calculation (dashed line) for (a), (c) and (e)
    $k_x$ and $k_y$ direction and (b), (d) and (f) $k_z$ direction.
    Here (a) and (b) is for $Bi_2Se_3$, (c) and (d) is for $Bi_2Te_3$,
    while (e) and (f) is for $Sb_2Te_3$. In (a), (c) and (e),
    the red line represents the dispersion along $k_x$ direction
    while the blue line is for $k_y$ direction.}
    \label{fig:dispersion8b}
\end{figure}

\section{New model Hamiltonian with eight bands}
\label{sec:8band}
As we have described above, our model Hamiltonian can capture the
salient topological features of the $Bi_2Se_3$ family of materials.
However, for full quantitative fitting with the first principle
calculations, we need to expand the basis set. By inspecting
carefully the ${\bf k}\cdot{\bf p}$ matrix elements, we find that
there are strong couplings between the state
$|P1^+_-,\pm\frac{1}{2}\rangle$ and the state
$|P2^-,\tilde{\Gamma}_{4,5}\rangle$ or
$|P2^-_-,\pm\frac{1}{2}\rangle$. For example, at the valence band
maximum $k_x\approx 0.07$\AA$^{-1}$, we find that these couplings
can be as large as the energy gap between these states. Therefore it
is not surprise that our model Hamiltonian with four bands is not
suitable in this regime. The strong couplings between these states
indicate that if we want to describe this material more accurately,
we need to further include the states
$|P2^-,\tilde{\Gamma}_{4,5}\rangle$ and
$|P2^-_-,\pm\frac{1}{2}\rangle$ into our model Hamiltonian. In the
basis sequence $|P1^+_-,\frac{1}{2}\rangle$,
$|P1^+_-,-\frac{1}{2}\rangle$, $|P2^-_+,\frac{1}{2}\rangle$,
$|P1^-_+,-\frac{1}{2}\rangle$, $|P2^-,\tilde\Gamma_4\rangle$,
$|P2^-,\tilde\Gamma_5\rangle$, $|P2^-_-,\frac{1}{2}\rangle$ and
$|P2^-_-,-\frac{1}{2}\rangle$, following the similar perturbation
procedure, we find that our model Hamiltonian is written as
\begin{widetext}
\begin{eqnarray}
    &&\hat{H}=\frac{\hbar^2}{2m_0}
    \left(
    \begin{array}{cccccccc}
        f_1({\bf k})&0&\frac{2}{\hbar}k_zQ_1&
        \frac{2}{\hbar}P_1k_-&\frac{2}{\hbar}Q_2k_+&\frac{2}{\hbar}k_+P_2&
        \frac{2}{\hbar}k_zQ_3&\frac{2}{\hbar}k_-P_3\\
        &f_1({\bf k})&\frac{2}{\hbar}k_+P_1^*&
        -\frac{2}{\hbar}k_zQ_1^*&-\frac{2}{\hbar}P_2^*k_-&\frac{2}{\hbar}Q_2^*k_-&
        \frac{2}{\hbar}P_3^*k_+&-\frac{2}{\hbar}Q_3^*k_z\\
        &&f_3({\bf k})&0&g_{35}({\bf k})&g_{36}({\bf k})
        &f_{37}({\bf k})&-g_{47}^*(-{\bf k})\\
        &&&f_3({\bf k})&g_{36}^*(-{\bf k})&-g_{35}^*(-{\bf k})
        &g_{47}({\bf k})&f_{37}^*(-{\bf k})\\&&&&f_5({\bf k})&0&
        -g_{68}^*(-{\bf k})&g_{58}({\bf k})\\&&h.c.&&&f_5({\bf k})&
        g_{58}^*(-{\bf k})&g_{68}({\bf k})\\&&&&&&f_{7}({\bf k})&0\\
        &&&&&&&f_7({\bf k})\\
    \end{array}
    \right)
\end{eqnarray}
\end{widetext}
with
\begin{eqnarray}
    f_{i(ij)}({\bf k})=F_{i(ij)}k_z^2+K_{i(ij)}k_{\parallel}^2\\
    g_{ij}({\bf k})=U_{ij}k_zk_++V_{ij}k_-^2.
\end{eqnarray}
The parameters $F_{ij}$, $K_{ij}$, $U_{ij}$ and $V_{ij}$ can now
also be determined by the perturbation theory, which is shown in
appendix \ref{app:C_kp}. In this Hamiltonian, time-reversal symmetry
is already satisfied. Furthermore, $R_2$ rotation symmetry yields
that $U_{35}=U_{36}^*$, $V_{35}=-V_{36}^*$, $U_{58}=-U_{68}^*$ and
$V_{58}=V_{68}^*$. The obtained parameters are listed in table
\ref{tab:8b_parameter} and the band dispersion is found to fit well
with that of {\it ab initio} calculation, as shown in Fig
\ref{fig:dispersion8b}. This demonstrate that the eight band model
is suitable to serve as the basis of the quantitative study of
$Bi_2Se_3$ family of materials.

We would like to make some more remarks about the eight band model.
Firstly, in our model Hamiltonian with four bands, the leading term
that breaks the in-plane full rotation symmetry down to $R_3$
symmetry is the third order in perturbation, while in the eight band
model, it is the second order coupling $g_{ij}$. These type of terms
exist because the states $|P2^-,\tilde\Gamma_{4,5}\rangle$
themselves break the rotation symmetry according to the expression
(\ref{eq:eff_Gamma4}) and (\ref{eq:eff_Gamma5}). Secondly, it is
interesting to compare the present eight band model with the
well-known Kane model for usual III-V or II-VI group semiconductors
with zinc-blend structure. In fact there is a one-to-one
correspondence between the basis of these two models, in which
$|P1^+_-,\pm\frac{1}{2}\rangle$ corresponds to the electron band
($\Gamma_6$), $|P2^-_+,\pm\frac{1}{2}\rangle$ and
$|P2^-,\Gamma_{4,5}\rangle$ correspond to the light hole and heavy
hole bands ($\Gamma_8$), respectively, and
$|P2^-_-,\pm\frac{1}{2}\rangle$ corresponds to the spin-orbit
split-off band ($\Gamma_7$). Therefore, from the symmetry point of
view, our model here is nothing but an extension of the Kane model
to a crystal structure with lower symmetry.

\section{Conclusions}
\label{sec:conclusion}
To summarize, based on the symmetry properties and the ${\bf
k}\cdot{\bf p}$ perturbation theory, we systematically derived a
model Hamiltonian for the 3D TI in the $Bi_2Se_3$ class of
materials. Our model Hamiltonian captures the main low energy
physics, such as the inverted band structure and topologically
protected surface states. The topological surface states have well
defined spin texture, which can be traced back to the sign of the
atomic SOC in these materials. Furthermore, the Landau levels of a
z-direction magnetic field for both bulk states and surface states
are calculated. The gap of bulk Landau levels is shown to decrease
when magnetic field increases, which may be observed in a
magneto-optical spectroscopy. Within the bulk gap, the surface
Landau levels appears as discrete peaks for the LDOS, which can be
detected by STM. We also analyze the quantitative limitation of our
model Hamiltonian with four bands and and show that a new model
Hamiltonian with eight energy bands can describe $Bi_2Se_3$ type of
materials quantitatively, which will be useful in the future
comparison with experiments.

\section{Acknowledgments}
We would like to thank Yulin Chen, Aahron Kapitulnik, Zhixun Shen and Qikun Xue for the helpful discussion. This work is supported by
the Department of Energy, Office of Basic Energy Sciences, Division
of Materials Sciences and Engineering, under contract
DE-AC02-76SF00515 and by the Keck Foundation. C.X. Liu acknowledge
financial support by the Alexander von Humboldt Foundation of Germany. This work is also supported by
the NSF of China, the National Basic Research Program of China (No. 2007CB925000), the International Science and Technology Cooperation Program of China (No. 2008DFB00170).


\begin{appendix}
\section{Symmetry property of group $D_{3d}^5$}
\label{app:A_symmetry}
As described in the text, the group $D_{3d}^5$ is generated by a
three-fold rotation operator $R_3$, a two-fold rotation operator
$R_2$ and an inversion operator $P$. It has six classes and
correspondingly six irreducable representations,
$\tilde{\Gamma}^\pm_1$, $\tilde{\Gamma}^\pm_2$ and
$\tilde{\Gamma}^\pm_3$ with the upper index $\pm$ denoting the
parity of the representation. Here we use $\tilde{\Gamma}$ to denote
the representation at $\Gamma$ point in BZ to avoid confusion with
the Dirac $\Gamma$ matrices. The character table of $D_{3d}^5$ is
given in table (\ref{tab:Character1})\cite{dresselhausbook2008}.
\begin{table}[htb]
  \centering
  \begin{minipage}[t]{0.8\linewidth} 
      \caption{ The character table for $D_{3d}^5(R\bar{3}m)$. }
\label{tab:Character1}
\hspace{-1cm}
\begin{tabular}{ccccccc}
    \hline\hline
    $D_{3d}(\bar{3}m)$  & $E$ & $2R_3$ & $3R_2$ & $P$ & $2PR_3$ & $3PR_2$ \\
    \hline
    $\tilde{\Gamma}_1^+$ &  1 &  1 &  1 &  1 &  1 &  1 \\
    $\tilde{\Gamma}_2^+$ &  1 &  1 & -1 &  1 &  1 & -1 \\
    $\tilde{\Gamma}_3^+$ &  2 & -1 &  0 &  2 & -1 &  0 \\
    $\tilde{\Gamma}_1^-$ &  1 &  1 &  1 & -1 & -1 & -1 \\
    $\tilde{\Gamma}_2^-$ &  1 &  1 & -1 & -1 & -1 &  1 \\
    $\tilde{\Gamma}_3^-$ &  2 & -1 &  0 & -2 &  1 &  0 \\
    \hline
  \end{tabular}%
  \end{minipage}
\end{table}

After taking into account the spin, the $\mathcal{C}=2\pi$ rotation
induces a minus sign for the spin part, so that the elements of the
group is doubled, which is the so-called double group. For
$D_{3d}^5$, the class and irreducable representation of the double
group is also doubled. The character table for the double group of
$D_{3d}^5$ is given in table
(\ref{tab:Character2})\cite{dresselhausbook2008}.

When constructing the double group, it is useful to consider the
decomposition of the direct product of $\tilde{\Gamma}_{1,2,3}^\pm$
and spinor represenation $\tilde\Gamma_6$, which is given by
\begin{eqnarray}
    &&\tilde\Gamma_3^\pm\otimes\tilde\Gamma_6^+=\tilde\Gamma_4^\pm+\tilde\Gamma_5^\pm+\tilde\Gamma_6^\pm
    \label{eq:App_Dirpro1}\\
    &&\tilde\Gamma_1^\pm\otimes\tilde\Gamma_6^+=\tilde\Gamma_6^\pm
    \label{eq:App_Dirpro2}\\
    &&\tilde\Gamma_2^\pm\otimes\tilde\Gamma_6^+=\tilde\Gamma_6^\pm.
        \label{eq:App_Dirpro3}
\end{eqnarray}

Furthermore when considering about the matrix elements of ${\bf
k}\cdot{\bf p}$ theory, the following direct productors will be
quite helpful.
\begin{eqnarray}
        &&(\tilde\Gamma_6^{\pm})^*\otimes\tilde\Gamma_6^\pm=\tilde\Gamma_1^++\tilde\Gamma_2^++\tilde\Gamma_3^+\label{eq:App_Dirpro4}\\
    &&(\tilde\Gamma_6^+)^*\otimes\tilde\Gamma_6^-=\tilde\Gamma_1^-+\tilde\Gamma_2^-+\tilde\Gamma_3^-\label{eq:App_Dirpro5}\\
    &&(\tilde\Gamma_6^+)^*\otimes\tilde\Gamma_4^\pm=\tilde\Gamma_3^\pm\label{eq:App_Dirpro6}\\
    &&(\tilde\Gamma_6^-)^*\otimes\tilde\Gamma_4^\pm=\tilde\Gamma_3^\mp\label{eq:App_Dirpro7}\\
    &&(\tilde\Gamma_6^+)^*\otimes\tilde\Gamma_5^\pm=\tilde\Gamma_3^\pm\label{eq:App_Dirpro8}\\
    &&(\tilde\Gamma_6^-)^*\otimes\tilde\Gamma_5^\pm=\tilde\Gamma_3^\mp\label{eq:App_Dirpro9}\\
    &&(\tilde\Gamma_{4(5)}^+)^*\otimes\tilde\Gamma_{4(5)}^-=\tilde\Gamma_1^-\label{eq:App_Dirpro10}\\
    &&(\tilde\Gamma_{4(5)}^+)^*\otimes\tilde\Gamma_{5(4)}^-=\tilde\Gamma_2^-\label{eq:App_Dirpro11}
\end{eqnarray}

\begin{widetext}
\begin{table}[h]
  \centering
  \begin{minipage}[t]{0.8\linewidth} 
      \caption{ The character table for the double group of $D_{3d}^5(R\bar{3}m)$. }
\label{tab:Character2} \hspace{0cm}
\begin{tabular}{ccccccccccccc}
    \hline\hline
    $D_{3d}(\bar{3}m)$  & $E$ & $2R_3$ & $3R_2$ & $P$ & $2PR_3$ & $3PR_2$
    & $\mathcal{C}$ & 2$\mathcal{C}R_3$ & $3\mathcal{C}R_2$ & $\mathcal{C}P$
    & $2\mathcal{C}PR_3$ & $3\mathcal{C}PR_2$\\
    \hline
    $\tilde{\Gamma_1}^+$ &  1 &  1 &  1 &  1 &  1 &  1 &  1 &  1 &  1 &  1 &  1 &  1\\
    $\tilde{\Gamma_2}^+$ &  1 &  1 & -1 &  1 &  1 & -1 &  1 &  1 & -1 &  1 &  1 & -1\\
    $\tilde{\Gamma_3}^+$ &  2 & -1 &  0 &  2 & -1 &  0 &  2 & -1 &  0 &  2 & -1 &  0\\
    $\tilde{\Gamma_4}^+$ &  1 & -1 &  i &  1 & -1 &  i & -1 &  1 & -i & -1 &  1 & -i\\
    $\tilde{\Gamma_5}^+$ &  1 & -1 & -i &  1 & -1 & -i & -1 &  1 &  i & -1 &  1 &  i\\
    $\tilde{\Gamma_6}^+$ &  2 &  1 &  0 &  2 &  1 &  0 & -2 & -1 &  0 & -2 & -1 &  0\\
    $\tilde{\Gamma_1}^-$ &  1 &  1 &  1 & -1 & -1 & -1 &  1 &  1 &  1 & -1 & -1 & -1\\
    $\tilde{\Gamma_2}^-$ &  1 &  1 & -1 & -1 & -1 &  1 &  1 &  1 & -1 & -1 & -1 &  1\\
    $\tilde{\Gamma_3}^-$ &  2 & -1 &  0 & -2 &  1 &  0 &  2 & -1 &  0 & -2 &  1 &  0\\
    $\tilde{\Gamma_4}^-$ &  1 & -1 &  i & -1 &  1 & -i & -1 &  1 & -i &  1 & -1 &  i\\
    $\tilde{\Gamma_5}^-$ &  1 & -1 & -i & -1 &  1 &  i & -1 &  1 &  i &  1 & -1 & -i\\
    $\tilde{\Gamma_6}^-$ &  2 &  1 &  0 & -2 & -1 &  0 & -2 & -1 &  0 &  2 &  1 &  0\\
   \hline
  \end{tabular}%
  \end{minipage}
\end{table}
\end{widetext}

\section{$\Gamma$ matrix}
\label{app:B_Gamma}

The five Dirac $\Gamma$ matrices can be defined as
\begin{eqnarray}
    &&\Gamma_1=\sigma_1\otimes\tau_1\qquad
    \Gamma_2=\sigma_2\otimes\tau_1\qquad
    \Gamma_3=\sigma_3\otimes\tau_1\nonumber\\
    &&\Gamma_4=1\otimes\tau_2\qquad
    \Gamma_5=1\otimes\tau_3,
    \label{GamMat_1}
\end{eqnarray}
which satifies Clifford algebra $\{\Gamma_a,\Gamma_b\}=2\delta_{ab}$.
The other ten $\Gamma$ matrices are given by $\Gamma_{ab}=[\Gamma_a,\Gamma_b]/2i$.
Explicitly, $\Gamma_{ab}$ is given by
\begin{eqnarray}
    &&\Gamma_{ij}=[\sigma_i\otimes\tau_1,\sigma_j\otimes\tau_1]/2i
    =\varepsilon_{ijk}\sigma_k\otimes1\\
    &&\Gamma_{i4}=[\sigma_i\otimes\tau_1,1\otimes\tau_2]/2i
    =\sigma_i\otimes\tau_3\\
    &&\Gamma_{i5}=[\sigma_i\otimes\tau_1,1\otimes\tau_3]/2i=-\sigma_i\otimes\tau_2\\
    &&\Gamma_{45}=[1\otimes\tau_2,1\otimes\tau_3]/2i=1\otimes\tau_1
    \label{GamMat_2}
\end{eqnarray}
where $i,j=1,2,3$.
Now let's check the properties of the fifteen $\Gamma$ matrice under
the time reversal operation $T$ and inversion operation $P$. We assume the
$\Gamma$ matrices are written in the basis $|P1^+_-,\frac{1}{2}\rangle$,
$|P2^-_+,\frac{1}{2}\rangle$, $|P1^+_-,-\frac{1}{2}\rangle$ and
$|P2^-_+,-\frac{1}{2}\rangle$, then the transformation matrix of
the symmetry operation has been obtained in Sec. \ref{sec:Eff}.
With these transformation matrices, we have
\begin{eqnarray}
    &&T\Gamma_iT^{-1}=P\Gamma_iP^{-1}=-\Gamma_i,\qquad i=1,2,3,4\\
    &&T\Gamma_5T^{-1}=P\Gamma_5P^{-1}=\Gamma_5
    \label{eq:Gamma5TP}
\end{eqnarray}
In fact $P$ operator is exactly $\Gamma_5$ here.
\begin{eqnarray}
    &&T\Gamma_{ij}T^{-1}=-P\Gamma_{ij}P^{-1}=-\Gamma_{ij},\\
    &&T\Gamma_{i4}T^{-1}=-P\Gamma_{i4}P^{-1}=-\Gamma_{i4},\\
    &&T\Gamma_{i5}T^{-1}=-P\Gamma_{i5}P^{-1}=\Gamma_{i5},\\
    &&T\Gamma_{45}T^{-1}=-P\Gamma_{45}P^{-1}=\Gamma_{45}.
    \label{eq:Gamma10TP}
\end{eqnarray}
where $i,j=1,2,3$.

Next let's consider about $R_2$.
\begin{eqnarray}
    &&R_2\Gamma_{1,4}R_2^{-1}=-\Gamma_{1,4},\\
    &&R_2\Gamma_{2,3,5}R_2^{-1}=\Gamma_{2,3,5}\\
    &&R_2\Gamma_{12,31,24,34,15,45}R_2^{-1}=-\Gamma_{12,31,24,34,15,45},\\
    &&R_2\Gamma_{23,14,25,35}R_2^{-1}=\Gamma_{23,14,25,35}.
    \label{eq:Gamma10R2}
\end{eqnarray}

Finally let's talk about the three fold rotation symmetry.
Under the rotation operation $R_z(\theta)$, the $\Gamma$
matrice are transformed as $\Gamma'(\theta)=
e^{i\frac{\Sigma}{2}\theta}\Gamma e^{-i\frac{\Sigma}{2}\theta}$, then
\begin{eqnarray}
    \frac{d\Gamma'(\theta)}{d\theta}=\frac{i}{2}[\Sigma,\Gamma'(\theta)]
    \label{eq:transform}
\end{eqnarray}
Therefore, the transformation properties of $\Gamma$ matrice under the rotation
operation are determined by the commutation relation $[\Sigma,\Gamma]$.
The commutation relations for $\Gamma$ matrice are listed as follows:
\begin{eqnarray}
    &&[\Sigma,\Gamma_1]=2i\Gamma_2,\qquad[\Sigma,\Gamma_2]=-2i\Gamma_1\\
    &&[\Sigma,\Gamma_3]=[\Sigma,\Gamma_4]=[\Sigma,\Gamma_5]=0,
    \nonumber\\
    &&[\Sigma,\Gamma_{12}]=0,\qquad [\Sigma,\Gamma_{34}]=0\\
    &&[\Sigma,\Gamma_{31}]=-2i\Gamma_{23}\qquad
    [\Sigma,\Gamma_{23}]=2i\Gamma_{31}\\
    &&[\Sigma,\Gamma_{14}]=2i\Gamma_{24}\qquad
        [\Sigma,\Gamma_{24}]=-2i\Gamma_{14}\\
    &&[\Sigma,\Gamma_{15}]=2i\Gamma_{25},\qquad
    [\Sigma,\Gamma_{25}]=-2i\Gamma_{15}\\
    &&[\Sigma,\Gamma_{35}]=0,\qquad [\Sigma,\Gamma_{45}]=0
    \label{eq:commutation}
\end{eqnarray}
With the above commutation relations, we can easily solve the equantion (\ref{eq:transform})
and find that
%
\begin{eqnarray}
    &&\Gamma'_1(\theta)=\Gamma_1\cos\theta-\Gamma_2\sin\theta,\nonumber\\
    &&\Gamma'_2(\theta)=\Gamma_1\sin\theta+\Gamma_2\cos\theta\\
    &&\Gamma'_3(\theta)=\Gamma_3,\qquad\Gamma'_4(\theta)=\Gamma_4\\
    &&\Gamma'_{23}(\theta)=\Gamma_{23}\cos\theta-\Gamma_{31}\sin\theta,\nonumber\\
    &&\Gamma'_{31}(\theta)=\Gamma_{31}\cos\theta+\Gamma_{23}\sin\theta\\
    &&\Gamma'_{14}(\theta)=\Gamma_{14}\cos\theta-\Gamma_{24}\sin\theta,\nonumber\\
    &&\Gamma'_{24}(\theta)=\Gamma_{14}\sin\theta+\Gamma_{24}\cos\theta\\
    &&\Gamma'_{15}(\theta)=\Gamma_{15}\cos\theta-\Gamma_{25}\sin\theta,\nonumber\\
    &&\Gamma'_{25}(\theta)=\Gamma_{15}\sin\theta+\Gamma_{25}\cos\theta\\
    &&\Gamma'_5(\theta)=\Gamma_5,\qquad\Gamma'_{34}=\Gamma_{34},
    \qquad\Gamma'_{12}=\Gamma_{12},\nonumber\\
    &&\Gamma'_{35}=\Gamma_{35},\qquad \Gamma'_{45}=\Gamma_{45}
    \label{eq:expGamma1}
\end{eqnarray}
The above results indicate that under the rotation $R_3$
$\Gamma_{3,4,5}$ and $\Gamma_{12,34,35,45}$ behave as scalars (or
pseudo-scalars), while the three pairs of operators
$\{\Gamma_{23},\Gamma_{31}\}$, $\{\Gamma_{14},\Gamma_{24}\}$ and
$\{\Gamma_{15},\Gamma_{25}\}$ behave as vectors. The corresponding
representation for each $\Gamma$ matrix is given in table
(\ref{tab:Character3}).

\begin{table}[htb]
  \centering
  \begin{minipage}[t]{0.8\linewidth} 
      \caption{ The character table of $\Gamma$ matrice and the polymals of
      the momentum ${\bf k}$.  }
\label{tab:Character3}
\hspace{0cm}
\begin{tabular}{ccc}
    \hline\hline
      &  Representation & T \\
    \hline
    $\{\Gamma_1,\Gamma_2\}$        & $\tilde{\Gamma}^-_3$  & - \\
    $\Gamma_3$                     & $\tilde{\Gamma}^-_1$  & - \\
    $\Gamma_4$                     & $\tilde{\Gamma}^-_2$  & - \\
    $\Gamma_5$                     & $\tilde{\Gamma}^+_1$  & + \\
    $\Gamma_{12}$                  & $\tilde{\Gamma}^+_2$  & - \\
    $\{\Gamma_{23},\Gamma_{31}\}$  & $\tilde{\Gamma}^+_3$  & - \\
    $\{\Gamma_{14},\Gamma_{24}\}$  & $\tilde{\Gamma}^+_3$  & - \\
    $\{\Gamma_{15},\Gamma_{25}\}$  & $\tilde{\Gamma}^-_3$  & + \\
    $\Gamma_{34}$                  & $\tilde{\Gamma}^+_2$  & - \\
    $\Gamma_{35}$                  & $\tilde{\Gamma}^-_1$  & + \\
    $\Gamma_{45}$                  & $\tilde{\Gamma}^-_2$  & + \\
    $\{k_x,k_y\}$                  & $\tilde{\Gamma}^-_3$  & - \\
    $k_z,k_z^3$                    & $\tilde{\Gamma}^-_2$  & - \\
    $1,k_x^2+k_y^2,k_z^2$          & $\tilde{\Gamma}^+_1$  & + \\
    $\{k_x^2-k_y^2,2k_xk_y\}$      & $\tilde{\Gamma}^+_3$  & + \\
    $k_x^3-3k_xk_y^2$              & $\tilde{\Gamma}^-_1$  & - \\
    $3k_x^2k_y-k_y^3$              & $\tilde{\Gamma}^-_2$  & - \\
    $\{k_x^3+k_xk_y^2,
    k_x^2k_y+k_y^3\}$              & $\tilde{\Gamma}^-_3$  & - \\
    $\{B_x,B_y\}$                  & $\tilde{\Gamma}^+_3$  & - \\
    $B_z$                          & $\tilde{\Gamma}^+_2$  & - \\
    \hline
  \end{tabular}%
  \end{minipage}
\end{table}

\section{Parameters in ${\bf k}\cdot{\bf p}$ theory}
\label{app:C_kp}

In this appendix, we hope to show the detail results from ${\bf
k}\cdot{\bf p}$ theory. First let's consider the constraint for the
matrix elements of the momentum from the $D_{3d}^5$ symmetry. As
described above, the eigen-states can be denoted by
$|\Lambda^\pm,\alpha\rangle$ with $\Lambda =P1_\pm,P2_\pm$ and
$\alpha=\pm\frac{1}{2},\pm\frac{3}{2}$. The states
$|\Lambda,\pm1/2\rangle$ belong to $\tilde{\Gamma}^{\pm}_6$
representation. For $|\Lambda,\pm3/2\rangle$, as described above, we
need to re-combine these two states as
\begin{eqnarray}
    &&|\Lambda^\pm,\tilde{\Gamma}_4\rangle=\frac{1}{\sqrt{2}}
    (|\Lambda^\pm,3/2\rangle+|\Lambda^\pm,-3/2\rangle)\\
    &&|\Lambda^\pm,\tilde{\Gamma}_5\rangle=\frac{1}{\sqrt{2}}
    (|\Lambda^\pm,3/2\rangle-|\Lambda^\pm,-3/2\rangle),
    \label{eq:app3_Gamma45}
\end{eqnarray}
which belong to $\tilde{\Gamma}_4$ and $\tilde{\Gamma}_5$
representation respectively. The expressions (\ref{eq:App_Dirpro4})
$\sim$ (\ref{eq:App_Dirpro11}) give the decomposition of the direct
product of these states. The momentum $p_x$, $p_y$ belongs to
$\tilde{\Gamma}^-_3$ representation, while $p_z$ belongs to
$\tilde{\Gamma}^-_2$ representation, therefore we require that the
decomposition of the direct product of the eigen-states also include
$\tilde{\Gamma}^-_3$ and $\tilde{\Gamma}^-_2$ to obtain non-zero
matrix elements. For example, the direct product of
$\tilde{\Gamma}^\pm_6$ and $\tilde{\Gamma}^\pm_{4,5}$ doesn't
contain $\tilde{\Gamma}_2$, which indicates that the matrix element
$\langle\Lambda_1,\pm1/2|p_z|\Lambda_2,\tilde{\Gamma}_{4,5} \rangle$
is always zero.

The symmetry operation can further help us to obtain
the relation between different matrix elements of the momentum.
For example, due to the $R_3$ rotation symmetry, we have
\begin{eqnarray}
&&\langle \Lambda_1^+,\frac{1}{2}|p_x|\Lambda_2^-,-\frac{1}{2}\rangle\nonumber\\
    &=&\langle \Lambda_1^+,\frac{1}{2}|R^\dag_3R_3p_xR^\dag_3R_3|
    \Lambda_2^-,-\frac{1}{2}\rangle\nonumber\\
    &=&e^{-i\frac{2\pi}{3}}\langle\Lambda^+_1,\frac{1}{2}|\left( p_x\cos\frac{2\pi}{3}
    -p_y\sin\frac{2\pi}{3}\right)|\Lambda^-_2,-\frac{1}{2}\rangle\nonumber\\
    &\rightarrow& \langle\Lambda_1^+,\frac{1}{2}|p_x|\Lambda_2^-,-\frac{1}{2}\rangle
    =i\langle\Lambda_1^+,\frac{1}{2}|p_y|\Lambda_2^-,-\frac{1}{2}\rangle.
    \label{eq:kp_eg}
\end{eqnarray}
Finally we can define the independent component of the matrix elements
as follows.
\begin{eqnarray}
    &&\langle \Lambda_1^+,\frac{1}{2}|p_x|\Lambda_2^-,-\frac{1}{2}\rangle
    =\langle \Lambda_1^+,-\frac{1}{2}|p_x|\Lambda_2^-,\frac{1}{2}\rangle\nonumber\\
        &&=i \langle\Lambda_1^+,\frac{1}{2}|p_y|\Lambda_2^-,-\frac{1}{2}\rangle=-i \langle
    \Lambda_1^+,-\frac{1}{2}|p_y|\Lambda_2^-,\frac{1}{2}\rangle\nonumber\\
    &&=P_{\Lambda_1^+,\Lambda_2^-}\\
        &&\langle \Lambda_1^+,\frac{1}{2}|p_z|\Lambda^-,\frac{1}{2}\rangle
    =-\langle \Lambda_1^+,-\frac{1}{2}|p_z|\Lambda^-,-\frac{1}{2}\rangle\nonumber\\
    &&=Q_{\Lambda_1^+,\Lambda_2^-}\\
    &&\langle \Lambda_1^\pm,\frac{1}{2}|p_x|\Lambda_2^\mp,\tilde\Gamma_4\rangle
    =-i \langle \Lambda_1^\pm,-\frac{1}{2}|p_x|\Lambda_2^\mp,\tilde\Gamma_4\rangle\nonumber\\
    &&=-i \langle \Lambda_1^\pm,\frac{1}{2}|p_y|\Lambda_2^\mp,\tilde\Gamma_4\rangle
    = \langle\Lambda_1^\pm,-\frac{1}{2}|p_y|\Lambda_2^\mp,\tilde\Gamma_4\rangle\nonumber\\
    &&=M_{\Lambda_1^\pm,\Lambda_2^\mp}\\
    &&\langle \Lambda_1^\pm,\frac{1}{2}|p_x|\Lambda_2^\mp,\tilde\Gamma_5\rangle
    =i \langle \Lambda_1^\pm,-\frac{1}{2}|p_x|\Lambda_2^\mp,\tilde\Gamma_5\rangle\nonumber\\
    &&=-i \langle \Lambda_1^\pm,\frac{1}{2}|p_y|\Lambda_2^\mp,\tilde\Gamma_5\rangle
        =-\langle\Lambda_1^\pm,-\frac{1}{2}|p_y|\Lambda_2^\mp,\tilde\Gamma_5\rangle\nonumber\\
    &&=N_{\Lambda_1^\pm,\Lambda_2^\mp}\\
    &&\langle \Lambda_1^+,\tilde\Gamma_4|p_z|\Lambda^-,\tilde\Gamma_4\rangle=R_{\Lambda_1^+,\Lambda_2^-}\\
    &&\langle \Lambda_1^+,\tilde\Gamma_5|p_z|\Lambda^-,\tilde\Gamma_5\rangle=S_{\Lambda_1^+,\Lambda_2^-}
    \label{eq:app3_para1}
\end{eqnarray}
Here it is more convenient of use $p_\pm=p_x\pm ip_y$, which leads
to
\begin{eqnarray}
    &&\langle \Lambda_1^+,\frac{1}{2}|p_+|\Lambda_2^-,-\frac{1}{2}\rangle
    =\langle \Lambda_1^+,-\frac{1}{2}|p_-|\Lambda_2^-,\frac{1}{2}\rangle \nonumber\\
    &&=2P_{\Lambda_1^+,\Lambda_2^-}\\
        &&\langle\Lambda_1^+,\frac{1}{2}|p_-|\Lambda_2^-,-\frac{1}{2}\rangle
    =\langle\Lambda_1^+,-\frac{1}{2}|p_+|\Lambda_2^-,\frac{1}{2}\rangle \nonumber\\
    &&=0\\
        &&\langle \Lambda_1^\pm,\frac{1}{2}|p_-|\Lambda_2^\mp,\Gamma_4\rangle
    =-i \langle \Lambda_1^\pm,-\frac{1}{2}|p_+|\Lambda_2^\mp,\Gamma_4\rangle \nonumber\\
    &&=2M_{\Lambda_1^\pm,\Lambda_2^\mp}\\
    &&\langle \Lambda_1^\pm,\frac{1}{2}|p_+|\Lambda_2^\mp,\Gamma_4\rangle
    =-i\langle\Lambda_1^\pm,-\frac{1}{2}|p_-|\Lambda_2^\mp,\Gamma_4\rangle\nonumber\\
    &&=0\\
    &&\langle \Lambda_1^\pm,\frac{1}{2}|p_-|\Lambda_2^\mp,\Gamma_5\rangle
    =i \langle \Lambda_1^\pm,-\frac{1}{2}|p_+|\Lambda_2^\mp,\Gamma_5\rangle\nonumber\\
    &&=2N_{\Lambda_1^\pm,\Lambda_2^\mp}\\
    && \langle \Lambda_1^\pm,\frac{1}{2}|p_+|\Lambda_2^\mp,\Gamma_5\rangle
    =i \langle\Lambda_1^\pm,-\frac{1}{2}|p_-|\Lambda_2^\mp,\Gamma_5\rangle\nonumber\\
    &&=0
    \label{eq:TE_para3}
\end{eqnarray}

Time reversal symmetry indicates that $P_{\Lambda_1^+,\Lambda_2^-}$,
$Q_{\Lambda_1^+,\Lambda_2^-}$ can be chosen to be real
($P_{\Lambda_1^+,\Lambda_2^-}=P^*_{\Lambda_1^+,\Lambda_2^-}$,
$Q_{\Lambda_1^+,\Lambda_2^-}=Q^*_{\Lambda_1^+,\Lambda_2^-}$) while
$M_{\Lambda_1^\pm,\Lambda_2^\mp}=iN^*_{\Lambda_1^\pm,\Lambda_2^\mp}$
and $R_{\Lambda_1^+,\Lambda_2^-}=-S^*_{\Lambda_1^+,\Lambda_2^-}$.
Since the matrix element between $|P1^+_-,\pm\frac{1}{2}\rangle$ and
$|P2^-_+,\pm\frac{1}{2}\rangle$ is quite important, we denote
\begin{eqnarray}
    &&\langle P1_-^+,\frac{1}{2}|p_x|P2_+^-,-\frac{1}{2}\rangle=\langle
    P1_-^+,-\frac{1}{2}|p_x|P2_+^-,\frac{1}{2}\rangle=\nonumber\\
        &&i \langle P1_-^+,\frac{1}{2}|p_y|P2^-_+,-\frac{1}{2}\rangle=-i \langle
    P1_-^+,-\frac{1}{2}|p_x|P2_+^-,\frac{1}{2}\rangle\nonumber\\
    &&=P_0\\
        &&\langle P1_-^+,\frac{1}{2}|p_z|P2_+^-,\frac{1}{2}\rangle
    =-\langle P1_-^+,-\frac{1}{2}|p_z|P2_+^-,-\frac{1}{2}\rangle\nonumber\\
    &&=Q_0
    \label{eq:TE_para2}
\end{eqnarray}

Now we consider the perturbation theory. The degenerate perturbation
formulism is given by
\begin{eqnarray}
    &&H^{(0)}_{mm'}=E_m\delta_{mm'}\label{eq:EH1_pert0}\\
    &&H^{(1)}_{mm'}=H'_{mm'}\label{eq:EH1_pert1}\\
    &&H^{(2)}_{mm'}=\frac{1}{2}\sum_{l}H'_{ml}H'_{lm'}
    \left( \frac{1}{E_m-E_l}\right.\nonumber\\
    &&\left.+\frac{1}{E_{m'}-E_l} \right)\label{eq:EH1_pert2}\\
    &&H^{(3)}_{mm'}=-\frac{1}{2}\sum_{l,m''}\left[ \frac{H'_{ml}H'_{lm''}H'_{m''m'}}
    {(E_{m'}-E_l)(E_{m''}-E_l)}\right.\nonumber\\
    &&\left.+\frac{H'_{mm''}H'_{m''l}H'_{lm'}}
    {(E_m-E_l)(E_{m''}-E_l)} \right]\nonumber\\
    &&+\frac{1}{2}\sum_{l,l'}H'_{ml}H'_{ll'}H'_{l'm'}
    \left[ \frac{1}{(E_m-E_l)(E_m-E_{l'})}\right.\nonumber\\
    &&\left.+\frac{1}{(E_{m'}-E_l)(E_{m'}-E_{l'})}
    \right].\label{eq:EH1_pert3}
\end{eqnarray}
Here $m$ and $m'$ are taken from $|P1^+_-,1/2\rangle=|1\rangle$,
$|P2^-_+,1/2\rangle=|2\rangle$, $|P1^+_-,-1/2\rangle=|3\rangle$ and
$|P2^-_+,-1/2\rangle=|4\rangle$ with the energy $E_1=E_3$ and
$E_2=E_4$ and $E_1<E_2$. $l$ is taken from the other bands except
for these four bands. The expression from perturbation calculation
of our model Hamiltonian with four bands is given as follows and the
values of the parameters are listed in table \ref{tab:4b_parameter}.
For our new model Hamiltonian with eight band model, the
perturbation procedure is the same to our model Hamiltonian with
four bands and here we only list the values of the parameters in
table \ref{tab:8b_parameter}.

\begin{eqnarray}
    &&C_0+M_0=E_1\label{eq:appC_para1}\\
    &&C_0-M_0=E_2\label{eq:appC_para2}\\
    &&C_1+M_1=\frac{\hbar^2}{2m_0}+\frac{\hbar^2}{m_0^2}
    \sum_{\Lambda^-}\frac{|Q_{P1^+,\Lambda^-}|^2}
    {E_{1}-E_{\Lambda^-,1/2}}\label{eq:appC_para3}\\
     &&C_2+M_2=\frac{\hbar^2}{2m_0}+\frac{\hbar^2}{m_0^2}
    \sum_{\Lambda^-}\left(\frac{|P_{P1^+,\Lambda^-}|^2}
    {E_{1}-E_{\Lambda^-,-1/2}}\right.\nonumber\\
    &&\left.+\frac{|M_{P1^+,\Lambda^-}|^2}
    {E_{1}-E_{\Lambda^-,\Gamma_4}}+\frac{|N_{P1^+,\Lambda^-}|^2}
    {E_{1}-E_{\Lambda^-,\Gamma_5}}\right) \label{eq:appC_para4}\\
    &&C_1-M_1=\frac{\hbar^2}{2m_0}+\frac{\hbar^2}{m_0^2}
    \sum_{\Lambda^-}\frac{|Q_{\Lambda^+,P2^-}|^2}
    {E_{1}-E_{\Lambda^+,1/2}}\label{eq:appC_para5}\\
    &&C_2-M_2=\frac{\hbar^2k^2}{2m_0}+\frac{\hbar^2}{m_0^2}
       \sum_{\Lambda^-}\left(\frac{|P_{\Lambda^+,P2^-}|^2}
       {E_{1}-E_{\Lambda^+,-1/2}}\right.\nonumber\\
       &&\left.+\frac{|M_{\Lambda^+,P2^-}|^2}
       {E_{1}-E_{\Lambda^+,\Gamma_4}}+\frac{|N_{\Lambda^+,P2^-}|^2}
       {E_{1}-E_{\Lambda^+,\Gamma_5}}\right)\label{eq:appC_para6}\\
    &&A_0=\frac{\hbar}{m_0}P_0\label{eq:appC_para7}\\
        &&B_0=\frac{\hbar}{m_0}Q_0\label{eq:appC_para8}
\end{eqnarray}

For $R_1$ and $R_2$ term we have
\begin{widetext}
\begin{eqnarray}
    &&R_1-R_2=\frac{\hbar^3}{m_0^3}\left[\sum_\Lambda
    \frac{|M_{P1^+,\Lambda^-}|^2P_{P1^+,P2^-}}{(E_{P2^-}-E_{\Lambda^-})
    (E_{P1^+}-E_{\Lambda^-})}-\right.\nonumber\\
    &&\left.\sum_{\Lambda_1^-,\Lambda^+_2}
    M_{P1^+\Lambda_1^-}M^*_{\Lambda_2^+\Lambda_1^-}P_{\Lambda_2^+P2^-}
    \left( \frac{1}{(E_{P1^+}-E_{\Lambda_1^-})(E_{P1^+}-E_{\Lambda_2^+})}
    +\frac{1}{(E_{P2^-}-E_{\Lambda_1^-})(E_{P2^-}-E_{\Lambda_2^+})}\right)
    \right]\label{eq:appC_para9}\\
    &&R_1+R_2=\frac{\hbar^3}{m_0^3}\left[-\sum_\Lambda
    \frac{P_{P1^+,P2^-}|M_{P2^-,\Lambda^-}|^2}{(E_{P2^-}-E_{\Lambda^+})
    (E_{P1^+}-E_{\Lambda^+})}-\right.\nonumber\\
    &&\left.\sum_{\Lambda_1^-,\Lambda^+_2}
    P_{P1^+\Lambda_1^-}M_{\Lambda_1^-\Lambda_2^+}M^*_{P2^-\Lambda_2^+}
    \left( \frac{1}{(E_{P1^+}-E_{\Lambda_1^-})(E_{P1^+}-E_{\Lambda_2^+})}
    +\frac{1}{(E_{P2^-}-E_{\Lambda_1^-})(E_{P2^-}-E_{\Lambda_2^+})}\right)
    \right]\label{eq:appC_para10}
\end{eqnarray}
\end{widetext}

\begin{table}[htb]
  \centering
  \begin{minipage}[t]{0.8\linewidth} 
      \caption{ The summary of the parameters in our model Hamiltonian with four
      bands. }
\label{tab:4b_parameter}
\begin{tabular}{cccc}
    \hline\hline
    & $Bi_2Se_3$ & $Bi_2Te_3$ & $Sb_2Te_3$ \\\hline
       $A_0 (eV\cdot$\AA) & 3.33 & 2.87 & 3.40 \\\hline
       $B_0 (eV\cdot$\AA) & 2.26 & 0.30 & 0.84 \\\hline
       $C_0 (eV)$ & -0.0083 & -0.18 & 0.001 \\\hline
       $C_1 (eV\cdot$\AA$^2)$ & 5.74 & 6.55 & -12.39 \\\hline
       $C_2 (eV\cdot$\AA$^2)$ & 30.4 & 49.68 & -10.78 \\\hline
       $M_0 (eV)$ & -0.28 & -0.30 & -0.22 \\\hline
       $M_1 (eV\cdot$\AA$^2)$ & 6.86 & 2.79 & 19.64 \\\hline
       $M_2 (eV\cdot$\AA$^2)$ & 44.5 & 57.38 & 48.51 \\\hline
       $R_1 (eV$\AA$^3)$ &  50.6 & 45.02 & 103.20 \\\hline
       $R_2 (eV$\AA$^3)$ &  -113.3 & -89.37 & -244.67 \\\hline
       $g_{1z}$ & -25.4 & -50.34 & -14.45 \\\hline
       $g_{1p}$ & -4.12 & -2.67 & -2.43 \\\hline
       $g_{2z}$ & 4.10 & 6.88 & 14.32 \\\hline
       $g_{2p}$ & 4.80 & 3.43 & 16.55 \\
       \hline\hline
  \end{tabular}%
\end{minipage}
\end{table}

Now we study the effect of magnetic field. Under magnetic field,
there are two different kinds of contribution. One is the orbital
term, which induce the Landau level and has been considered in
Sec.~\ref{sec:LL}. The other one is the Zeeman type term, which is
described by an effective g factor. In the following we will discuss
about the effective g factor in detail. There are two kinds of
contributions to the effective g factor. One comes from the atomic g
factor, which can be estimated from the {\it ab initio} calculation.
In an atom, the electron spin and orbital angular momentum couples
to magnetic field by $\hat{H}_{Zee}=\frac{\mu_B}{\hbar} \left(
g_l{\bf L}+g_s{\bf S}\right)\cdot{\bf B}=\frac{\mu_B}{\hbar}g_0 {\bf
J}\cdot{\bf B}$, where ${\bf J}={\bf S}+{\bf L}$ is the total
angular momentum and $g_0$ is so called Land$e$ g-factor. The wave
functions for the basis of our model Hamiltonian have been
calculated from {\it ab initio} calculation, which can be projected
into the atomic orbitals. Since for each atomic orbitals, the g
factor is simply given by
$g_0=1+\frac{J(J+1)-L(L+1)+S(S+1)}{2J(J+1)}$, the effective $g_0$
can be easily calculated, which is found to be $g_0\approx 1.2$.
Another contribution to the effective g factor origins from the
second order perturbation, which is related to the correction to the
effective mass term. The relation between the effective mass and
effective g factor in the ordinary semiconductors is known as the
Roth's formula\cite{roth1959}. Here the second order correction to g
factor is given by
\begin{eqnarray}
    &&g_{1z}^{(2)}=\frac{4}{m_0}
    \sum_{\Lambda^-,\alpha}\left(
    \frac{|P_{P1^+,\Lambda^-}|^2}{E_1-E_{\Lambda^-,-1/2}}\right.\nonumber\\
    &&\left.-\frac{|M_{P1^+,\Lambda^-}|^2}
    {E_1-E_{\Lambda^-,\Gamma_4}}-\frac{|N_{P1^+,\Lambda^-}|^2}
    {E_1-E_{\Lambda^-,\Gamma_5}}\right)\\
    &&g^{(2)}_{1p}=\frac{4}{m_0}\sum_{\Lambda^-}\frac{
    Q_{P1^+,\Lambda^-}P^*_{P1^+,\Lambda^-}}{E_{P1^+}-E_{\Lambda^-,1/2}}\\
    &&g_{2z}^{(2)}=\frac{4}{m_0}
    \sum_{\Lambda^+,\alpha}\left(\frac{|P_{\Lambda^+,P2^-}|^2}
    {E_{2}-E_{\Lambda^+,-1/2}}\right.\nonumber\\
    &&\left.-\frac{|M_{\Lambda^+,P2^-}|^2}
    {E_{2}-E_{\Lambda^+,\Gamma_4}}-\frac{|N_{\Lambda^+,P2^-}|^2}
    {E_{2}-E_{\Lambda^+,\Gamma_5}}\right)\\
    &&g^{(2)}_{2p}=\frac{4}{m_0}\sum_{\Lambda^+}\frac{
    Q^*_{\Lambda^+,P2^-}P_{\Lambda^+,P2^-}}{E_{P2^-}-E_{\Lambda^+}}.
    \label{eq:appC_geff2}
\end{eqnarray}
where $g_{1z(p)}$ and $g_{2z(p)}$ are defined in (\ref{eq:kp_Hmag2}).
Therefore finally our effective g factor is the summation of the
above two different contributions,
\begin{eqnarray}
    &&g_{\alpha}=g_0+g^{(2)}_\alpha, \qquad \alpha=1z,2z,1p,2p
    \label{eq:appC_geff}
\end{eqnarray}
and the values of effective g factor are given in table \ref{tab:4b_parameter}.
From table (\ref{tab:4b_parameter}), we find that
for $|P1^+_-,\pm\frac{1}{2}\rangle$ band, there is a strong
anisotropy which comes from the large contribution of the second order
perturbation of the states $|P2^-,\pm\frac{3}{2}\rangle$ and
$|P2^-_-,\pm\frac{1}{2}\rangle$.

\begin{widetext}
\begin{table}[htb]
  \centering
  \begin{minipage}[t]{0.8\linewidth} 
      \caption{ The summary of the parameters in the eight band
      effective model. }
\label{tab:8b_parameter}
\begin{tabular}{cccc}
    \hline\hline
     & $Bi_2Se_3$ & $Bi_2Te_3$ & $Sb_2Te_3$ \\\hline
       $P_1 (eV\cdot$\AA) & 3.33 & 2.87 & 3.40 \\\hline
       $Q_1 (eV\cdot$\AA) & 2.26 & 0.30 & 0.84 \\\hline
       $P_2 (eV\cdot$\AA) & 2.84 & 2.68 & 3.19 \\\hline
       $Q_2 (eV\cdot$\AA) & 2.84 & 2.68 & 3.19 \\\hline
       $P_3 (eV\cdot$\AA) & -2.62 & -1.94 & -2.46 \\\hline
       $Q_3 (eV\cdot$\AA) & 2.48 & 1.23 & 2.11 \\\hline
       $F_1 (eV\cdot$\AA$^2)$ & 3.73 & 7.16 & 3.82 \\\hline
       $K_1 (eV\cdot$\AA$^2)$ & 6.52 & 3.72 & 2.49 \\\hline
       $F_3 (eV\cdot$\AA$^2)$ & -1.12 & 3.76 & -32.03 \\\hline
       $K_3 (eV\cdot$\AA$^2)$ & -14.0 & -7.70 & -59.28 \\\hline
       $F_5 (eV\cdot$\AA$^2)$ & 1.50 & -0.62 & -2.26 \\\hline
       $K_5 (eV\cdot$\AA$^2)$ & -3.11 & -7.17 & -13.00 \\\hline
       $F_7 (eV\cdot$\AA$^2)$ & 2.71 & 3.77 & 5.04 \\\hline
       $K_7 (eV\cdot$\AA$^2)$ & -5.08 & 22.27 & 2.40 \\\hline
       $U_{35}=U_{36}^* (eV\cdot$\AA$^2)$ & -2.31-7.45i & -2.21-9.85i & -11.31-46.00i  \\\hline
       $V_{35}=-V_{36}^* (eV\cdot$\AA$^2)$ & -1.05-5.98i & -2.43-3.53i & -4.50-22.80i \\\hline
       $F_{37} (eV\cdot$\AA$^2)$ & 2.47 & 4.39 & 16.96 \\\hline
       $K_{37} (eV\cdot$\AA$^2)$ & -8.52 & -6.50 & -24.17 \\\hline
       $U_{47} (eV\cdot$\AA$^2)$ & -7.86 & -4.29 & -45.46 \\\hline
       $V_{47} (eV\cdot$\AA$^2)$ & -8.95i & -0.83i & -17.64i \\\hline
       $U_{58}=-U_{68}^* (eV\cdot$\AA$^2)$ & -2.31-2.57i & -0.24-3.69i & -2.01-3.98i \\\hline
       $V_{58}=V_{68}^* (eV\cdot$\AA$^2)$ & -0.64-4.29i & -0.85-6.64i & 1.28-9.02i \\\hline
       $E_1 (eV)$ & -0.29 & -0.48 & -0.22 \\\hline
       $E_3 (eV)$ & 0.28 & 0.12 & 0.22 \\\hline
       $E_5 (eV)$ & -0.57 & -0.63 & -0.88 \\\hline
       $E_7 (eV)$ & -0.98 & -1.18 & -1.51 \\
      \hline\hline
  \end{tabular}%
\end{minipage}
\end{table}
\end{widetext}

\end{appendix}

%

\end{document}